\documentclass[twocolumn]{aastex7}

\usepackage{graphicx}
\usepackage{amsmath}
\usepackage[encapsulated]{CJK}
\usepackage{ucs}
\usepackage{chngcntr}
\usepackage{xcolor}
\graphicspath{{./}{figs/}}

\newcommand{\msae}{\texttt{msaexp}}
\newcommand{\unite}{\texttt{unite}}
\newcommand{\cloudy}{\texttt{Cloudy}}
\newcommand{\pyneb}{\texttt{pyneb}}
\newcommand{\sirocco}{\texttt{Sirocco}}

\newcommand{\kms}{\rm km\,s^{-1}}

\newcommand{\dv}{\Delta v}

\newcommand{\zspec}{z_{\rm spec}}

\newcommand{\ha}{H$\alpha$}
\newcommand{\hb}{H$\beta$}
\newcommand{\hg}{H$\gamma$}
\newcommand{\hd}{H$\delta$}

\newcommand{\oii}{[O\,{\sc ii}]}
\newcommand{\oiii}{[O\,{\sc iii}]}
\newcommand{\nii}{[N\,{\sc ii}]}
\newcommand{\sii}{[S\,{\sc ii}]}
\newcommand{\neiii}{[Ne\,{\sc iii}]}
\newcommand{\hei}{He\,\textsc{i}}
\newcommand{\heii}{He\,\textsc{ii}}

\newcommand{\neiiiw}{[Ne\,{\sc iii}]\,$\lambda$3870}
\newcommand{\oiiidoubb}{\oiii\,$\lambda$5008}
\newcommand{\heib}{He\,\textsc{i}\,$\lambda$5877}

\newcommand{\oiiidoub}{\oiii\,$\lambda\lambda$4960,5008}
\newcommand{\niidoub}{\nii\,$\lambda\lambda$6550,6585}
\newcommand{\siidoub}{\sii\,$\lambda\lambda$4070,4078}
\newcommand{\oiiiauroral}{\oiii\,$\lambda$4364}
\newcommand{\oiin}{\oii\,$\lambda$3727}

\newcommand{\feii}{[Fe\,{\sc ii}]}

\newcommand{\heiir}{He\,$\textsc{ii}\,\lambda$4687}

\newcommand{\ham}{{\rm H}\alpha}
\newcommand{\hbm}{{\rm H}\beta}
\newcommand{\hgm}{{\rm H}\gamma}

\newcommand{\caii}{Ca\,{\sc ii}}
\newcommand{\caiik}{\caii\,K\,$\lambda$3935}

\newcommand{\necrit}{n_{\rm crit}}

\newcommand{\vabs}{v_{\rm abs}}
\newcommand{\fwhm}{{\rm FWHM}}

\newcommand{\vtermb}{v_{\rm term}^{\rm blue}}
\newcommand{\vtermr}{v_{\rm term}^{\rm red}}

\newcommand{\rd}{RUBIES-EGS-49140}
\newcommand{\ms}{UNCOVER-A2744-45924}
\newcommand{\fe}{FRESCO-GN-9771}

\newcommand{\qso}{A2744-QSO1}

\shorttitle{Balmer Absorption Series}
\shortauthors{Wang et al.}

\begin{document}
\begin{CJK*}{UTF8}{gbsn}

\title{Balmer Absorption Series and Broad Metal Lines in Two Luminous Little Red Dots}

\correspondingauthor{Bingjie Wang}
\email{bjwang@princeton.edu}

\author[0000-0001-9269-5046]{Bingjie Wang (王冰洁)}
\thanks{NHFP Hubble Fellow}
\affiliation{Department of Astrophysical Sciences, Princeton University, Princeton, NJ 08544, USA}
\email{bjwang@princeton.edu}

\author[0000-0002-5612-3427]{Jenny E. Greene}
\affiliation{Department of Astrophysical Sciences, Princeton University, Princeton, NJ 08544, USA}
\email{jgreene@astro.princeton.edu}

\author[0000-0003-2488-4667]{Hanpu Liu (刘翰溥)}
\affiliation{Department of Astrophysical Sciences, Princeton University, Princeton, NJ 08544, USA}
\email{hanpu.liu@princeton.edu}

\author[0000-0002-5375-8232]{Nicholas Kaaz}
\affiliation{Princeton Center for Theoretical Science, Princeton University, Princeton, NJ 08544, USA}
\affiliation{Princeton Gravity Initiative, Princeton University, Princeton, NJ 08544, USA}
\email{nkaaz@princeton.edu}

\author[0000-0003-2680-005X]{Gabriel B. Brammer}
\affiliation{Cosmic Dawn Center (DAWN), Copenhagen, Denmark}
\email{gabriel.brammer@nbi.ku.dk}

\author[0000-0002-4684-9005]{Raphael E. Hviding}
\affiliation{Max-Planck-Institut f\"ur Astronomie, D-69117 Heidelberg, Germany}
\email{hviding@mpia.de}

\author[0000-0002-2057-5376]{Ivo Labb\'e}
\affiliation{Centre for Astrophysics and Supercomputing, Swinburne University of Technology, Melbourne, VIC 3122, Australia}
\email{ilabbe@swin.edu.au}

\author[0000-0001-6755-1315]{Joel Leja}
\affiliation{Department of Astronomy \& Astrophysics, The Pennsylvania State University, University Park, PA 16802, USA}
\affiliation{Institute for Computational \& Data Sciences, The Pennsylvania State University, University Park, PA 16802, USA}
\affiliation{Institute for Gravitation and the Cosmos, The Pennsylvania State University, University Park, PA 16802, USA}
\email{joel.leja@psu.edu}

\author[0000-0003-2871-127X]{Jorryt Matthee}
\affiliation{Institute of Science and Technology Austria (ISTA), 3400 Klosterneuburg, Austria}
\email{jorryt.matthee@ista.ac.at}

\author[0000-0003-3997-5705]{Rohan P. Naidu}
\affiliation{Institute for Astronomy, University of Hawai`i, Honolulu, HI 96822, USA}
\email{rnaidu@hawaii.edu}

\author[0000-0001-5586-6950]{Alberto Torralba}
\affiliation{Institute of Science and Technology Austria (ISTA), 3400 Klosterneuburg, Austria}
\email{alberto.torralba@ista.ac.at}

\author[0009-0005-2295-7246]{Josephine F.W. Baggen}
\affiliation{Department of Astronomy, Yale University, New Haven, CT 06511, USA}
\email{josephine.baggen@yale.edu}

\author[0000-0001-7151-009X]{Nikko J. Cleri}
\affiliation{Department of Astronomy \& Astrophysics, The Pennsylvania State University, University Park, PA 16802, USA}
\affiliation{Institute for Computational \& Data Sciences, The Pennsylvania State University, University Park, PA 16802, USA}
\affiliation{Institute for Gravitation and the Cosmos, The Pennsylvania State University, University Park, PA 16802, USA}
\email{cleri@psu.edu}

\author[0000-0001-7201-5066]{Seiji Fujimoto}
\affiliation{David A. Dunlap Department of Astronomy and Astrophysics, University of Toronto, Toronto, ON M5S 3H4, Canada}
\affiliation{Dunlap Institute for Astronomy and Astrophysics, Toronto, ON M5S 3H4, Canada}
\email{seiji.fujimoto@utoronto.ca}

\author[0000-0001-6278-032X]{Lukas J. Furtak}
\affiliation{Department of Astronomy, The University of Texas at Austin, Austin, TX 78712, USA}
\affiliation{Cosmic Frontier Center, The University of Texas at Austin, Austin, TX 78712, USA}
\email{furtak@post.bgu.ac.il}

\author[0000-0002-2380-9801]{Anna de Graaff}
\affiliation{Max-Planck-Institut f\"ur Astronomie, D-69117 Heidelberg, Germany}
\email{degraaff@mpia.de}

\author[0000-0002-3301-3321]{Michaela Hirschmann}
\affiliation{Institute of Physics, Laboratory for Galaxy Evolution, EPFL, Observatory of Sauverny, 1290 Versoix, Switzerland}
\email{michaela.hirschmann@epfl.ch}

\author[0000-0002-5588-9156]{Vasily Kokorev}
\affiliation{Department of Astronomy, The University of Texas at Austin, Austin, TX 78712, USA}
\affiliation{Cosmic Frontier Center, The University of Texas at Austin, Austin, TX 78712, USA}
\email{vasily.kokorev.astro@gmail.com}

\author[orcid=0000-0003-3216-7190,sname='Lambrides']{Erini Lambrides}
\affiliation{Astrophysics Science Division, NASA Goddard Space Flight Center, Greenbelt, MD 20771, USA}
\affiliation{Department of Astronomy, University of Maryland, College Park, MD 20742, USA}
\affiliation{Center for Research and Exploration in Space Science and Technology, NASA/GSFC, Greenbelt, MD 20771, USA}
\email{erini.lambrides@nasa.gov}  

\author[0000-0002-2446-8770]{Ian McConachie}
\affiliation{Department of Astronomy, University of Wisconsin-Madison, Madison, WI 53706, USA}
\email{ian.mcconachie@wisc.edu}

\author[0000-0002-7524-374X]{Erica J. Nelson}
\affiliation{Department for Astrophysical \& Planetary Science, University of Colorado, Boulder, CO 80309, USA}
\email{erica.june.nelson@colorado.edu}

\author[0000-0003-0390-0656]{Ad\`ele Plat}
\affiliation{Institute of Physics, Laboratory for Galaxy Evolution, EPFL, Observatory of Sauverny, 1290 Versoix, Switzerland}
\email{adele.plat@epfl.ch}

\author[0000-0002-9593-8274]{Weichen Wang}
\affiliation{Dipartimento di Fisica G. Occhialini, Universit\`a degli Studi di Milano-Bicocca, I-20126 Milano, Italy}
\email{weichen.wang@unimib.it}

\author[0000-0002-0350-4488]{Adi Zitrin}
\affiliation{Department of Physics, Ben-Gurion University of the Negev, Be'er-Sheva 84105, Israel}
\email{adizitrin@gmail.com}

\begin{abstract}

Balmer absorption is common among little red dots (LRDs), but absorbers at or redward of systemic are rare, occurring in only $\sim10-15$\% of H$\alpha$ absorbers. In this paper, we study two such exceptional cases with deep JWST/NIRSpec spectroscopy: $15$~hr of high-resolution (G395H) observations of RUBIES-EGS-49140 ($\zspec=6.68$), resolving the absorption in all four transitions from H$\alpha$ through H$\delta$, and medium-resolution spectroscopy ($10$~hr of G235M, $2$~hr of G395M) of UNCOVER-A2744-45924 ($\zspec=4.46$; $1.7\times$ magnification). Both sources are among the optically reddest and most luminous LRDs known, and both show deep, near-systemic Balmer absorption troughs. We find two systematic trends along the Balmer series: the absorption centroids become more redshifted toward higher-order transitions, while the absorbed equivalent widths decline only weakly with increasing order, far less than expected from the atomic optical-depth ratios for a single attenuating screen. Ca\,{\sc{ii}}\,K is detected in absorption in both sources, whose offset follows the H$\alpha$ trough rather than the more redshifted higher-order Balmer lines. We further report the detection of a broad base in [Ne\,{\sc{iii}}]\,$\lambda$3870, along with broad [O\,{\sc{iii}}]\,$\lambda$4364, [O\,{\sc{iii}}]\,$\lambda5008$, and He\,{\sc{i}}\,$\lambda5877,\lambda7067$, while He\,{\sc{ii}}\,$\lambda$4687 remains undetected or weak. Standard AGN photoionization models cannot reproduce the observed line ratios, whereas AGNs with high gas densities provide a consistent explanation, as also indicated by the anomalously high He\,{\sc{i}}\,$\lambda7067/\lambda5877$ ratio. A possible explanation for the relative strengths of the Balmer absorption lines could be a dense, optically thick medium whose re-emission modifies their apparent absorption strengths, while the velocity progression may arise from stratification in the absorbing gas.

\end{abstract}


\section{Introduction\label{sec:intro}}

A population of compact, luminous red sources discovered with the James Webb Space Telescope (JWST), commonly referred to as Little Red Dots (LRDs; \citealt{Matthee2024}), has emerged as one of the most intriguing discoveries of recent years. First identified as compact sources with unusually red spectral energy distributions (SEDs) (e.g., \citealt{Barro2024, Kokorev2024, Labbe2023:agn, Akins2025:cw}), LRDs have since been the subject of extensive spectroscopic follow-up.
Their spectra reveal the near-ubiquity of broad Balmer emission, pointing to an active galactic nucleus (AGN) origin \citep[e.g.,][]{Greene2024, Maiolino2024, Hviding2025, Kocevski2025, Taylor2025, Zhuang2026}. Yet LRDs exhibit a distinctive combination of properties that challenges conventional models of AGN and their host galaxies. Among their most distinctive spectroscopic features is strong Balmer absorption \citep[e.g.,][]{Matthee2024, DEugenio2025:fe, Torralba2026:fe}.

Balmer absorption requires a large column of hydrogen with substantial population in the $n=2$ level along the line of sight. Before JWST, such absorption was known only in a handful of rare, low-ionization broad absorption-line (LoBAL) quasars, where the absorption troughs typically span thousands of $\kms$ \citep{Hall2002, Aoki2010, Schulze2018}. Strong, but much narrower, Balmer absorption is now reported in $\sim$40\% of LRDs, with the true incidence potentially higher because such features can be missed at low spectral resolution or signal to noise (S/N); by comparison, the incidence is only $\sim$0.05\% rate found in low-redshift type~1 AGNs \citep{Shangguan2026:sdss, Yanagisawa2026:atlas}.
Together with the discovery of LRDs exhibiting Balmer breaks stronger than any stellar population can produce \citep{deGraaff2025:cliff, Naidu2025}, these observations point to a dense neutral gas component that has become a key piece in several models of the population \citep[e.g.,][]{Begelman2026, Inayoshi2025:dense, Kido2025, Liu2025, Nandal2026}.

Many LRDs show prominent P-Cygni profiles in their Balmer lines that are interpreted as signatures of outflowing gas, while a small number of LRDs instead show absorption close to the systemic velocity \citep{Chen2026:abcd, Matthee2026, Juodzbalis2026:census}, raising the possibility of a qualitatively different geometry or kinematic state.
The current census remains limited, however. 
First, most existing LRD spectra are either obtained with the low-resolution Prism or with relatively shallow medium-resolution grating observations. At $R\sim1000$, the absorption is strongly resolution-limited, making it difficult to constrain the intrinsic strength and kinematics of the absorber.
Second, most studies focus on \ha\ and/or \hb\ \citep{Matthee2026, Juodzbalis2026:census} (or, at lower redshift, on the Paschen-line dynamics; e.g., \citealt{Juodzbalis2024:rs, Wang2025:brd, Kokorev2026:glimpse}), while few have studied through \hg\ or higher-order Balmer transitions \citep[e.g.,][]{Chen2026:abcd, DEugenio2025:fe}.
Yet the much fainter higher-order lines provide a crucial lever on the physical structure of the absorber, as systematic trends across the Balmer series can reveal its kinematic and optical-depth structure.
The underlying physics is straightforward: the line-center optical depth of a Balmer transition scales as $\tau_0 \propto n_2\,f\,\lambda$, where $n_2$ is the column density of hydrogen in the $n=2$ level, $f$ is the oscillator strength of the transition, and $\lambda$ is its rest wavelength. The oscillator strength drops steeply along the series ($f=0.641$, 0.119, 0.0447, and 0.0221 for \ha\ through \hd), so each higher order transition is intrinsically less opaque and therefore thermalizes at deeper radii in the absorbing gas.

These limitations leave several basic questions open. Do the higher-order Balmer lines exhibit systematic trends that depart from the null hypothesis of a single, homogeneous medium of neutral gas? This null hypothesis makes two predictions. First, the absorption centroid should be common to every transition, since all four troughs arise from the same gas.
Second, the line-center optical depths should follow the atomic ratios. The absorbed equivalent widths (EWs) should therefore decline steeply along the series, unless all four transitions are saturated, in which case the troughs instead share a common depth set by covering factor, $C_f$, alone.
What physical properties, then, distinguish the rare near-systemic absorbers from the more common blueshifted, P-Cygni population? Helium and metal lines offer an independent window into the physical conditions of the emitting gas, but detecting and deblending these faint features requires deep, medium- to high-resolution spectroscopy.

In this work, we present 15~hr of high-resolution JWST/NIRSpec G395H spectroscopy of \rd\ ($z=6.68$), resolving the Balmer absorption in all four transitions from \ha\ through \hd. We also analyze a medium-resolution spectrum of \ms\ ($z=4.46$), one of the most luminous LRDs known.

The two sources offer complementary advantages.
\rd\ was among the earliest LRDs identified with a strong Balmer break in JWST/NIRSpec Prism spectroscopy \citep{Wang2024:ub}.
Subsequent studies have identified a rich suite of Fe emission lines \citep{DEugenio2025:fe, Lambrides2025:fevii}, reported possible spectral variability \citep{Lambrides2026} (though \citealt{Liu2026:var} found a flat light curve with different temporal sampling), and detected broad \oiiiauroral\ emission in deep medium-resolution spectroscopy \citep{Papovich2026}.
\ms\ \citep{Labbe2024:monster}, albeit observed at lower spectral resolution, has exceptionally high S/N, and its high luminosity also makes typically faint emission robustly measurable in both its narrow and broad components. MUSE observations have in addition revealed N\,\textsc{iv}]\,$\lambda$1486 and Ly$\alpha$ emission \citep{Torralba2026:monster}. Together, these datasets provide the resolution and depth needed to characterize the Balmer absorption across four transitions and the faint metal-line emission and its associated physical conditions.

The structure of this paper is as follows.
Section~\ref{sec:data} introduces the dataset.
Section~\ref{sec:method} describes the line-decomposition methodology.
Section~\ref{sec:res} presents the decomposition and characterization of \ha--\hd, the identification of broad helium and metal lines, most notably \neiiiw, and the photoionization diagnostics.
Section~\ref{sec:dis} discusses physical interpretations of the observed trends in absorber velocities and depths across the Balmer series.
We conclude in Section~\ref{sec:concl}.

A flat $\Lambda$CDM cosmology with $H_{0}=70$ ${\rm km \,s^{-1} \,Mpc^{-1}}$, $\Omega_{M}=0.3$, and $\Omega_{\Lambda}=0.7$ is assumed.
Where applicable, velocities are quoted relative to the narrow component of \oiiidoubb, and, unless otherwise stated, we report the posterior median with uncertainties given by the 16th and 84th percentiles.

\section{Data\label{sec:data}}

\subsection{\rd\label{sec:data:rd}}

\begin{figure*}
\gridline{
  \fig{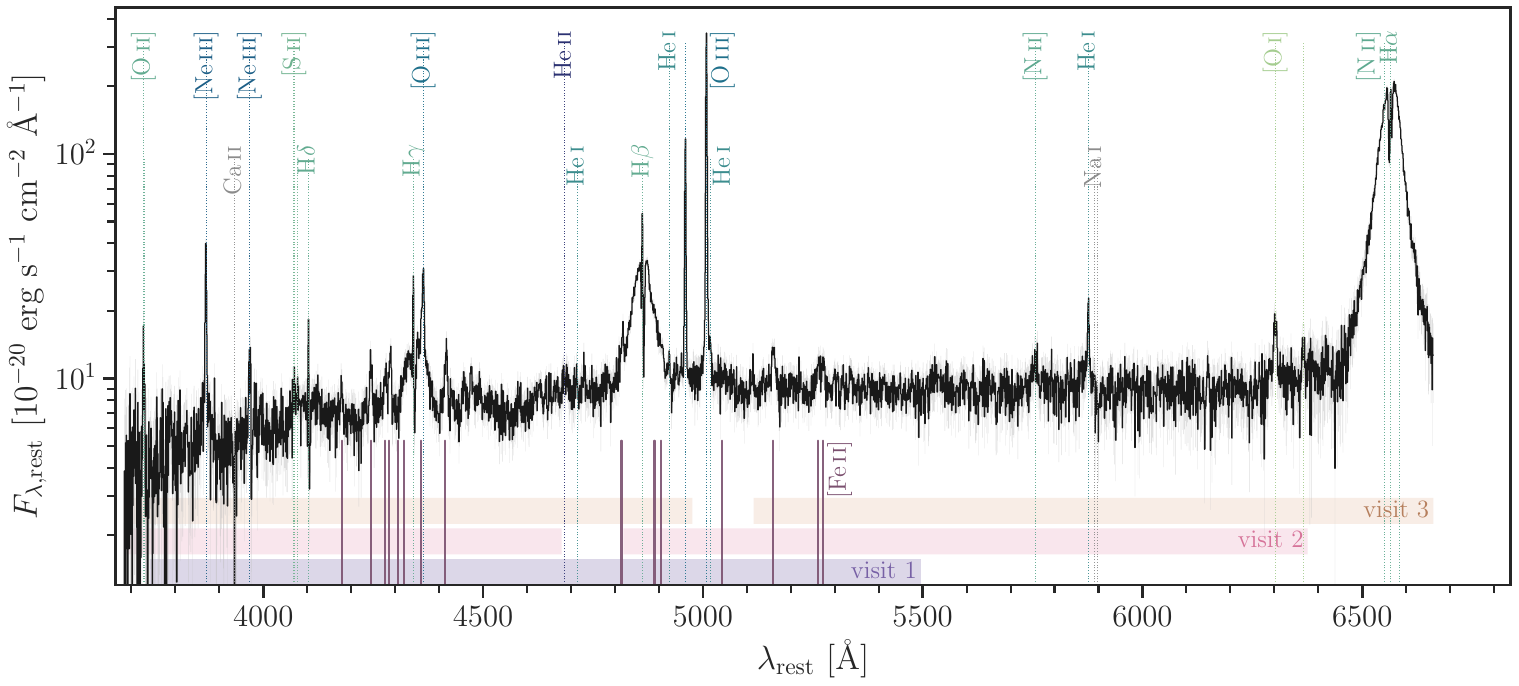}{0.95\textwidth}{}
}
\caption{G395H/F290LP spectrum of \rd\ ($z=6.68$), combining three 5~hr visits whose
  wavelength coverage is annotated. The emission lines are labeled, and the blue vertical lines indicate the prominent \feii\ emission lines.
}
\label{fig:data_rd}
\end{figure*}

\rd\ ($\zspec=6.68$) is an LRD spectroscopically identified in the RUBIES survey \citep{deGraaff2025:rubies, Wang2024:ub}.
It was observed in the G395H/F290LP mode as part of JWST-GO-8047 (PIs Wang \& Nelson). To cover all four Balmer lines (\ha--\hd), the observation was split over three visits of 5~hr each.
The source was placed on different slitlets in each visit, so that its spectrum fell on different parts of the detector. In all three visits, we manually positioned the source at the center of the slitlet.
Figure~\ref{fig:data_rd} shows the stacked total spectrum, with the spectral coverage of each visit annotated; zoom-ins on the Balmer lines are shown in Figure~\ref{fig:fit_balmer}.
We note that this stacked spectrum is not used in the actual kinematic modeling. Instead, as detailed later in Section~\ref{sec:method}, the three spectra are fit simultaneously, with terms that account for their relative flux and wavelength calibrations. This ensures that the measured velocity offsets are not driven by calibration differences between the individual spectra.

These high-resolution spectra are supplemented with an archival 7.7~hr medium-resolution G395M/F290LP spectrum from JWST-GO-4106 (PIs Nelson \& Labb\'e; \citealt{DEugenio2025:fe}).
We note that for a point source the effective resolution is typically higher than the nominal slit-filled values. Because the source does not fill the slit, an idealized point source yields $R \approx 3600$--$5400$ ($\dv \approx 56$--$83~\kms$) for G395H \citep{deGraaff2024}; in practice, however, the effective resolution is lower.
All spectra are reduced with \msae, corresponding to version 4 on the DAWN JWST Archive\footnote{\url{https://dawn-cph.github.io/dja}} \citep{Brammer2022, deGraaff2024, Heintz2024}.

\subsection{\ms\label{sec:data:ms}}

\ms\ ($\zspec=4.46$) is an extremely luminous ($L_{\rm H\alpha} \approx 10^{44}~{\rm erg\,s^{-1}}$) LRD identified in the UNCOVER survey \citep{Bezanson2024, Labbe2024:monster}, with the \ha\ spectrum taken from All the Little Things  (ALT; \citealt{Naidu2024:alt}). 
It is magnified by $\mu=1.7\pm0.2$, based on the UNCOVER strong lens model of A2744 \citep{Furtak2023:lens}, updated with UNCOVER NIRSpec and ALT grism redshifts \citep{Price2025}. The luminosities and fluxes reported herein are corrected for this magnification factor.

\begin{figure*}
\gridline{
  \fig{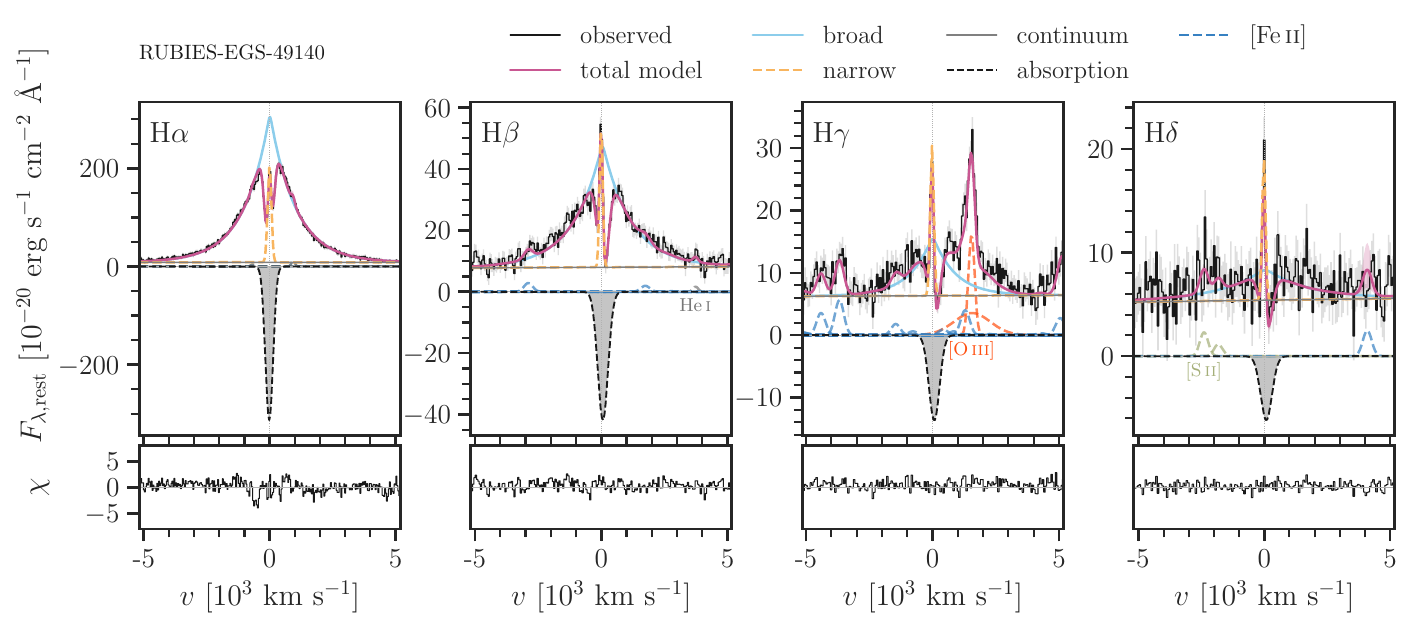}{0.95\textwidth}{(a)}
}
\gridline{
  \fig{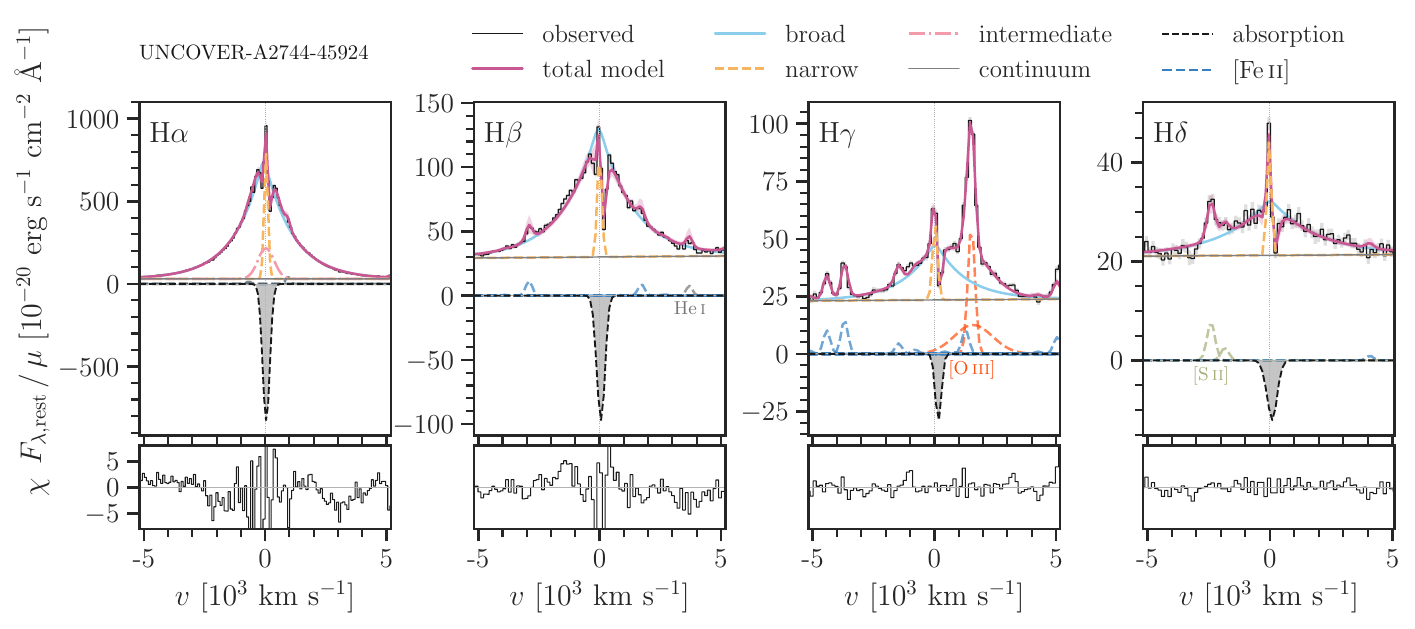}{0.95\textwidth}{(b)}
}
\caption{Model fits to the Balmer lines of (a) \rd\ and (b) \ms.
  In each panel the data (black) are shown with the total model and its
  components: narrow Gaussian emission, broad exponential emission,
  Gaussian absorption, and, for \ha\ in \ms, an intermediate-width Gaussian emission.
  Residuals are shown below each panel.
}
\label{fig:fit_balmer}
\end{figure*}

It was more recently observed in G235M/F170LP for 9.7~hr and G395M/F290LP for 2.4~hr as part of JWST-GO-8204 (PIs Greene \& Labb\'e), and reduced in the same way with \msae. \hb\ and the higher-order Balmer lines fall in G235M ($R \approx 1100$--$2300$, $\dv \approx 130$--$273~\kms$ for an idealized point source), and \ha\ in G395M ($R \approx 1300$--$2000$, $\dv \approx 150$--$231~\kms$).
Although the resolution is lower than \rd, the exceptional S/N, with a median S/N per pixel of $\approx 25$ across the full spectra and $\approx 70$ in the \ha\ window, makes this source a valuable complement. Zoom-ins on the four Balmer lines are likewise shown in Figure~\ref{fig:fit_balmer}.

\section{Measurements\label{sec:method}}

\subsection{Line Profile Fitting\label{sec:method:fit}}

All lines are fit simultaneously using \unite\ \citep{unite}, a JAX-based package that models multiple spectra jointly and accounts for the under-sampling of the NIRSpec line spread function (LSF) by integrating the model within each pixel. We adopt the NIRSpec LSF for an idealized point source \citep{deGraaff2024}, with a nuisance parameter that rescales the LSF width to absorb the associated systematic uncertainty. Additional nuisance terms capture the relative flux calibration and wavelength calibration of each spectrum.

Each Balmer line is modeled with three components: a narrow Gaussian emission component, a broad exponential emission component, and a Gaussian absorption component. The only exception is \ha\ in \ms, where previous studies have found that an additional intermediate-width Gaussian is required \citep[e.g.,][]{Matthee2026}, and we follow the same approach. The components are not tied across transitions. Instead, we impose dependent priors requiring the narrow-emission fluxes to follow the ordering \oiii\ $>$ \hb\ $>$ \hg\ $>$ \hd. This prior is intended to break the degeneracy between narrow-emission infilling and the depth of the absorption trough by preventing the narrow-emission component from trading freely against the absorber. A similar approach was explored for \rd\ using medium-resolution spectroscopy \citep{DEugenio2025:fe}, and our results are consistent within $1\sigma$.

The remaining identified lines within the covered wavelength range (\oiin,  \neiiiw, \siidoub, \oiiiauroral, \oiiidoub, \heib, [O\,{\sc i}]\,$\lambda\lambda$6302,6366, and \hei\,$\lambda$7067) are each modeled with a Gaussian profile, with doublet ratios fixed to their atomic values where applicable. The \feii\ multiplets are fit simultaneously as Gaussian components with shared kinematics. Although \niidoub\ is expected to be very weak, as found in e.g., \citealt{Greene2024, Hviding2025}, we include it in the model nevertheless. For lines that exhibit a broad base (\heib, $\lambda$7067 in Section~\ref{sec:res:hei}; \neiiiw, \oiiiauroral\ and \oiiidoub\ in Section~\ref{sec:res:metal}), we include an additional broad component.
As some of these broad components are reported here for the first time, we additionally test both exponential profiles for the broad emission and compare the resulting fits.

Our decomposition is chosen to be the simplest model that adequately reproduces the observed spectra. Despite the exceptional quality of the data, degeneracies remain among the spectral components, and multiple parameterizations can potentially reproduce the observed line profiles equally well. Distinguishing between alternative decompositions will likely require detailed full radiative transfer models.

\subsection{Terminal Velocities of Absorbers\label{sec:method:vterm}}

Approximating the absorption with a Gaussian provides a useful empirical description in the absence of a detailed physical model for the absorber. However, the intrinsic absorption profile need not be Gaussian, so the fitted Gaussian centroid may not robustly trace the absorber kinematics.

We therefore additionally characterize each absorption trough by its terminal velocities---the blue- and red-side edges where the absorber transmission returns toward the continuum---following \citet{Torralba2026:panbh}. From each posterior sample we construct the transmission profile, $T(v) = 1 - |A(v)|/[B(v)+C]$,
where $A$ is the Gaussian absorption component, $B$ the broad emission component, and $C$ the continuum. The narrow emission components are assumed to lie in front of the absorber and so are excluded. The blue- and red-side terminal velocities, $\vtermb$ and $\vtermr$, are defined as the outermost velocities at which $T$ crosses $0.95$, evaluated for each posterior sample.
We adopt the $95\%$ transmission level \citep{Naidu2026:sn}, rather than $99\%$, since the latter can be unstable. In our case, however, the two definitions yield the same results.

Unlike the Gaussian centroid, the terminal velocities are largely insensitive to the detailed shape of the absorption profile. For a single absorbing layer with the same kinematic extent across all transitions, one expects $\vtermb$ and $\vtermr$ to be identical for every Balmer line. Significant differences between transitions would instead indicate that the absorption is probing different effective optical depths, or more generally that the absorber cannot be described by a single homogeneous component. The primary systematic uncertainty arises from blending with neighboring \hei\ and \neiii\ emission, which is accounted for in our joint model.

\section{Results\label{sec:res}}

The decomposition and characterization of \ha--\hd, together with the identification of broad lines, most notably in \neiii, comprise the two major findings of this paper. In this section, we first present the hydrogen and helium profiles (Section~\ref{sec:res:balmer}-\ref{sec:res:he}). We then turn to the metal lines (Section~\ref{sec:res:metal}) and the photoionization diagnostics enabled by these measurements (Section~\ref{sec:res:ratios}).

\subsection{The Balmer Series of \ha--\hd\label{sec:res:balmer}}

\subsubsection{Near-Systemic Emission and Absorption}

\begin{figure*}
\gridline{
  \fig{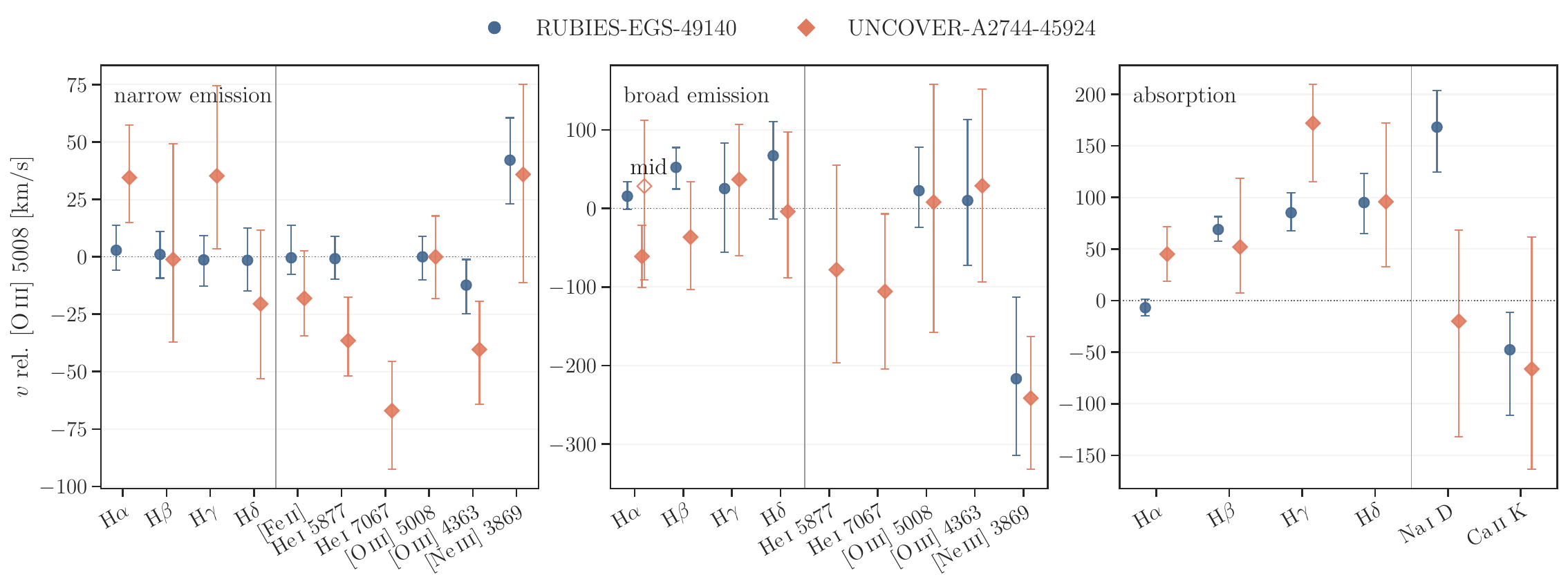}{0.95\textwidth}{(a)}
}
\gridline{
  \fig{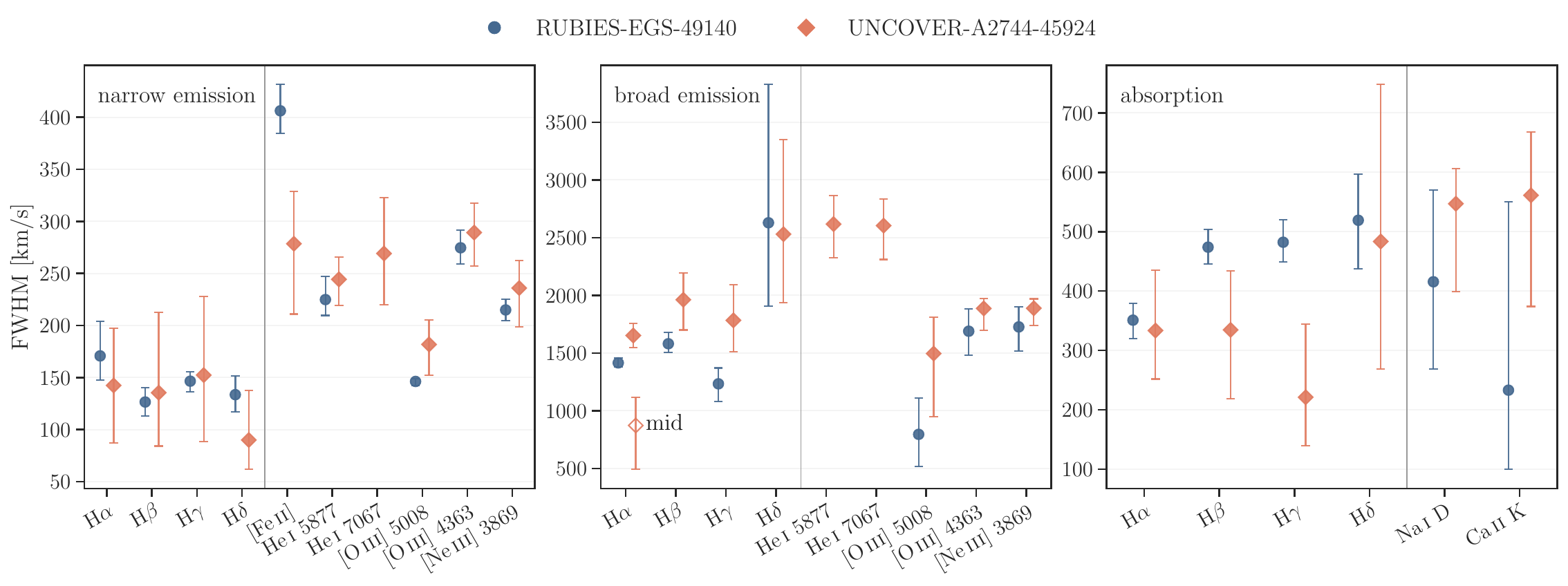}{0.95\textwidth}{(b)}
}
\caption{Model component summaries across all the lines examined for
  \rd\ (blue circles) and \ms\ (orange diamonds): (a) velocity offsets relative to narrow \oiiidoubb,
  and (b) FWHM. Each panel shows the narrow emission, broad emission, or
  absorption components. The intermediate-width \ha\ component required for \ms\ is shown as an unfilled diamond.
  The absorbers sit within $\lesssim100$--$200~\kms$ of systemic in every transition, with a mild tendency to be more redshifted in the higher-order Balmer lines.
}
\label{fig:components}
\end{figure*}

\begin{deluxetable*}{llrrrrrrrr}
\tabletypesize{\footnotesize}
\tablecaption{Line Decomposition\label{tab:balmer}}
\tablehead{
  \colhead{} & \colhead{Component}
  & \multicolumn{4}{c}{\rd\ ($z=6.68$)}
  & \multicolumn{4}{c}{\ms\ ($z=4.46$)} \\
  \cline{3-6}\cline{7-10}
  \colhead{} & \colhead{}
  & \colhead{$v$\tablenotemark{a}} & \colhead{FWHM} & \colhead{$F$} & \colhead{EW\tablenotemark{b}}
  & \colhead{$v$\tablenotemark{a}} & \colhead{FWHM} & \colhead{$F$\tablenotemark{c}} & \colhead{EW\tablenotemark{b}} \\
  \colhead{} & \colhead{}
  & \colhead{($\kms$)} & \colhead{($\kms$)}
  & \colhead{($10^{-18}\,{\rm erg}$} & \colhead{(\AA)}
  & \colhead{($\kms$)} & \colhead{($\kms$)}
  & \colhead{($10^{-18}\,{\rm erg}$} & \colhead{(\AA)} \\
  \colhead{} & \colhead{} & \colhead{} & \colhead{}
  & \colhead{${\rm s^{-1}\,cm^{-2}})$} & \colhead{}
  & \colhead{} & \colhead{}
  & \colhead{${\rm s^{-1}\,cm^{-2}})$} & \colhead{}
}
\startdata
\ha & narrow & $3^{+11}_{-9}$ & $171^{+33}_{-23}$ & $8.42^{+6.03}_{-2.17}$ & $82.0^{+58.7}_{-21.2}$ & $35^{+23}_{-19}$ & $142^{+55}_{-55}$ & $36.4^{+47.7}_{-11.9}$ & $109^{+142}_{-35}$ \\
 & intermediate & \nodata & \nodata & \nodata & \nodata & $28^{+84}_{-119}$ & $872^{+246}_{-379}$ & $45.4^{+46.0}_{-27.3}$ & $136^{+137}_{-82}$ \\
 & broad & $15^{+19}_{-16}$ & $1415^{+45}_{-35}$ & $145^{+5}_{-5}$ & $1415^{+46}_{-49}$ & $-61^{+40}_{-40}$ & $1652^{+105}_{-103}$ & $414^{+25}_{-23}$ & $1236^{+74}_{-67}$ \\
 & absorption & $-8^{+11}_{-9}$ & $351^{+29}_{-31}$ & $-28.2^{+2.8}_{-4.9}$ & $-9.3^{+0.9}_{-1.6}$ & $45^{+29}_{-31}$ & $333^{+102}_{-82}$ & $-81.1^{+32.1}_{-73.7}$ & $-9.1^{+3.6}_{-8.3}$ \\
\hb & narrow & $1^{+10}_{-10}$ & $126^{+14}_{-13}$ & $1.29^{+0.17}_{-0.17}$ & $13.1^{+1.7}_{-1.7}$ & $-1^{+51}_{-36}$ & $135^{+77}_{-51}$ & $3.72^{+2.66}_{-1.60}$ & $11.2^{+8.0}_{-4.8}$ \\
 & broad & $52^{+25}_{-27}$ & $1580^{+98}_{-77}$ & $16.3^{+0.6}_{-0.6}$ & $165^{+6}_{-6}$ & $-37^{+71}_{-67}$ & $1962^{+234}_{-262}$ & $52.8^{+4.1}_{-4.3}$ & $159^{+12}_{-13}$ \\
 & absorption & $71^{+14}_{-15}$ & $474^{+30}_{-28}$ & $-3.78^{+0.28}_{-0.30}$ & $-7.9^{+0.6}_{-0.6}$ & $54^{+65}_{-46}$ & $335^{+100}_{-116}$ & $-7.59^{+2.74}_{-3.25}$ & $-5.3^{+1.9}_{-2.3}$ \\
\hg & narrow & $-1^{+11}_{-11}$ & $147^{+9}_{-10}$ & $0.69^{+0.06}_{-0.06}$ & $8.8^{+0.8}_{-0.8}$ & $35^{+39}_{-32}$ & $152^{+76}_{-64}$ & $1.40^{+0.64}_{-0.39}$ & $5.4^{+2.5}_{-1.5}$ \\
 & broad & $25^{+58}_{-81}$ & $1233^{+137}_{-156}$ & $2.75^{+0.34}_{-0.27}$ & $35.0^{+4.4}_{-3.4}$ & $37^{+70}_{-97}$ & $1783^{+306}_{-274}$ & $10.2^{+1.7}_{-1.4}$ & $39.3^{+6.5}_{-5.4}$ \\
 & absorption & $82^{+20}_{-21}$ & $482^{+38}_{-34}$ & $-1.16^{+0.16}_{-0.17}$ & $-6.6^{+0.9}_{-0.9}$ & $175^{+38}_{-60}$ & $221^{+123}_{-82}$ & $-1.56^{+0.50}_{-0.57}$ & $-2.7^{+0.9}_{-1.0}$ \\
\hd & narrow & $-2^{+14}_{-13}$ & $134^{+18}_{-17}$ & $0.35^{+0.06}_{-0.04}$ & $5.3^{+0.8}_{-0.6}$ & $-21^{+32}_{-33}$ & $90^{+47}_{-28}$ & $0.82^{+0.36}_{-0.20}$ & $3.5^{+1.5}_{-0.9}$ \\
 & broad & $67^{+43}_{-81}$ & $2629^{+1198}_{-723}$ & $1.65^{+0.52}_{-0.29}$ & $24.6^{+7.8}_{-4.3}$ & $-4^{+101}_{-84}$ & $2530^{+822}_{-593}$ & $6.56^{+1.30}_{-1.27}$ & $27.9^{+5.5}_{-5.4}$ \\
 & absorption & $98^{+32}_{-30}$ & $519^{+77}_{-82}$ & $-0.52^{+0.09}_{-0.10}$ & $-5.3^{+1.0}_{-1.0}$ & $94^{+66}_{-63}$ & $483^{+264}_{-214}$ & $-1.09^{+0.43}_{-0.44}$ & $-2.8^{+1.1}_{-1.2}$ \\
\hline
\oiii\,$\lambda$5008 & narrow & $0^{+9}_{-10}$ & $146^{+3}_{-4}$ & $10.3^{+0.4}_{-0.4}$ & $99.7^{+3.4}_{-4.3}$ & $0^{+18}_{-18}$ & $182^{+24}_{-30}$ & $33.8^{+1.9}_{-2.1}$ & $98.1^{+5.6}_{-6.1}$ \\
 & broad & $23^{+55}_{-46}$ & $795^{+316}_{-280}$ & $1.45^{+0.29}_{-0.21}$ & $14.2^{+2.8}_{-2.1}$ & $8^{+150}_{-166}$ & $1496^{+313}_{-547}$ & $7.16^{+2.58}_{-2.56}$ & $20.8^{+7.5}_{-7.4}$ \\
\oiiiauroral & narrow & $-12^{+11}_{-13}$ & $275^{+17}_{-16}$ & $0.79^{+0.05}_{-0.04}$ & $10.0^{+0.6}_{-0.6}$ & $-40^{+21}_{-24}$ & $289^{+28}_{-32}$ & $3.28^{+0.24}_{-0.25}$ & $12.6^{+0.9}_{-1.0}$ \\
 & broad & $10^{+103}_{-83}$ & $1689^{+195}_{-209}$ & $1.03^{+0.17}_{-0.20}$ & $13.1^{+2.2}_{-2.5}$ & $29^{+123}_{-122}$ & $1887^{+83}_{-191}$ & $3.92^{+0.84}_{-0.62}$ & $15.1^{+3.2}_{-2.4}$ \\
\neiiiw & narrow & $-5^{+12}_{-14}$ & $215^{+10}_{-10}$ & $1.09^{+0.05}_{-0.05}$ & $19.8^{+1.0}_{-0.9}$ & $-35^{+25}_{-22}$ & $236^{+26}_{-37}$ & $4.27^{+0.19}_{-0.25}$ & $22.6^{+1.0}_{-1.3}$ \\
 & broad & $-217^{+104}_{-98}$ & $1726^{+172}_{-210}$ & $0.67^{+0.09}_{-0.10}$ & $12.2^{+1.7}_{-1.8}$ & $-242^{+78}_{-90}$ & $1888^{+82}_{-151}$ & $3.81^{+0.39}_{-0.35}$ & $20.2^{+2.1}_{-1.9}$ \\
\oii\,$\lambda$3727 & narrow & $42^{+18}_{-19}$ & $389^{+32}_{-39}$ & $0.37^{+0.08}_{-0.07}$ & $8.2^{+1.8}_{-1.6}$ & $36^{+39}_{-47}$ & $209^{+100}_{-98}$ & $0.62^{+0.16}_{-0.15}$ & $5.2^{+1.4}_{-1.3}$ \\
\oii\,$\lambda$3730 & narrow & $42^{+18}_{-19}$ & $389^{+32}_{-39}$ & $0.05^{+0.06}_{-0.04}$ & $1.2^{+1.4}_{-0.9}$ & $36^{+39}_{-47}$ & $209^{+100}_{-98}$ & $0.21^{+0.17}_{-0.13}$ & $1.7^{+1.4}_{-1.1}$ \\
\heii\,$\lambda$4686\tablenotemark{d} & narrow & $0^{+9}_{-10}$ & $480^{+14}_{-30}$ & $0.17^{+0.03}_{-0.03}$ & $1.8^{+0.3}_{-0.3}$ & $0^{+18}_{-18}$ & $455^{+31}_{-65}$ & $0.36^{+0.06}_{-0.06}$ & $1.1^{+0.2}_{-0.2}$ \\
\heib & narrow & $-1^{+10}_{-9}$ & $225^{+22}_{-15}$ & $0.62^{+0.03}_{-0.04}$ & $5.8^{+0.3}_{-0.4}$ & $-36^{+19}_{-15}$ & $244^{+21}_{-25}$ & $2.23^{+0.14}_{-0.18}$ & $6.3^{+0.4}_{-0.5}$ \\
 & broad & \nodata & \nodata & \nodata & \nodata & $-78^{+133}_{-119}$ & $2617^{+245}_{-291}$ & $3.75^{+0.58}_{-0.55}$ & $10.5^{+1.6}_{-1.5}$ \\
\hei\,$\lambda$7067 & narrow & \nodata & \nodata & \nodata & \nodata & $-67^{+21}_{-25}$ & $269^{+54}_{-49}$ & $2.54^{+0.34}_{-0.31}$ & $7.9^{+1.1}_{-1.0}$ \\
 & broad & \nodata & \nodata & \nodata & \nodata & $-106^{+99}_{-98}$ & $2604^{+233}_{-294}$ & $5.58^{+0.62}_{-0.72}$ & $17.5^{+1.9}_{-2.3}$ \\
\feii\tablenotemark{e} & multiplet & $-0^{+14}_{-7}$ & $406^{+25}_{-22}$ & \nodata & \nodata & $-18^{+21}_{-16}$ & $279^{+50}_{-68}$ & \nodata & \nodata \\
\hline
\caii\,$\lambda$3934 & absorption & $-46^{+39}_{-41}$ & $224^{+328}_{-122}$ & $-0.10^{+0.02}_{-0.05}$ & $-1.7^{+0.4}_{-0.8}$ & $-58^{+134}_{-97}$ & $573^{+91}_{-180}$ & $-0.39^{+0.13}_{-0.18}$ & $-1.8^{+0.6}_{-0.8}$ \\
Na\,{\sc i}\,$\lambda$5892\tablenotemark{f} & absorption & $118^{+75}_{-109}$ & $480^{+143}_{-169}$ & $-0.11^{+0.06}_{-0.06}$ & $-1.0^{+0.6}_{-0.6}$ & $-13^{+83}_{-115}$ & $542^{+66}_{-146}$ & $-0.26^{+0.15}_{-0.16}$ & $-0.73^{+0.41}_{-0.45}$ \\
Na\,{\sc i}\,$\lambda$5898\tablenotemark{f} & absorption & $118^{+75}_{-109}$ & $480^{+143}_{-169}$ & $-0.11^{+0.06}_{-0.08}$ & $-1.1^{+0.5}_{-0.7}$ & $-13^{+83}_{-115}$ & $542^{+66}_{-146}$ & $-0.18^{+0.08}_{-0.10}$ & $-0.49^{+0.22}_{-0.28}$ \\
\enddata
\tablenotetext{a}{Velocities are relative to narrow \oiiidoubb.}
\tablenotetext{b}{For the Balmer absorption, the EWs are measured against a differently defined continuum: the total model with the narrow (and, for \ms\ \ha, intermediate) emission and the absorber itself removed.}
\tablenotetext{c}{Fluxes are intrinsic, i.e., divided by the lensing magnification $\mu=1.7$.}
\tablenotetext{d}{Redshift is fixed to that of narrow \oiiidoubb.}
\tablenotetext{e}{The \feii\ multiplet is fit as a single kinematic suite, so only $v$ and FWHM are reported.}
\tablenotetext{f}{The Na\,{\sc i} doublet is fit with a shared redshift and FWHM.}
\end{deluxetable*}

In both sources, we resolve the narrow component and detect Balmer absorption in all four transitions. 
As summarized in Figure~\ref{fig:components} and also Table~\ref{tab:balmer},
the narrow emission, broad emission, and absorption components are all centered close to the systemic velocity. 

In \rd, the \ha\ absorber is consistent with the systemic velocity within $1\sigma$ ($\vabs = -8^{+11}_{-9}~\kms$), while the higher-order absorbers are mildly redshifted ($+71^{+14}_{-15}$, $+82^{+20}_{-21}$, and $+98^{+32}_{-30}~\kms$ for \hb, \hg, and \hd). 
The same qualitative pattern holds in \ms, albeit with larger uncertainties.
The velocity pattern previously noted in \rd\ (\ha\ blueshifted, \hb--\hd\ redshifted; cf. \citealt{DEugenio2025:fe}) thus also emerges from our spectral decomposition. Notably, neither object exhibits the blueshifted P-Cygni profiles seen in many other LRDs \citep{Matthee2026}.

The narrow emission components have typical widths of $\fwhm\approx90$--$180~\kms$, while the broad emission components have $\fwhm\approx1230$--$1960~\kms$. The absorbers are in-between, with $\fwhm\approx200$--$530~\kms$. In \rd, both the broad-emission and absorption widths increase mildly toward higher-order transitions (e.g., the absorber broadens slightly from $351^{+29}_{-31}~\kms$ at \ha\ to $482^{+38}_{-34}~\kms$ at \hg); in \ms\ the absorber widths are consistent within $1$--$2\sigma$. Although the \hd\ broad component is broader in both objects ($\fwhm\approx2500$--$2700~\kms$), its lower S/N makes the measurement uncertain, and we do not place significant weight on this apparent trend with Balmer transition.

\subsubsection{Terminal Velocities\label{sec:res:vterm}}

The velocity pattern described above is based on the centroids of Gaussian absorbers. As an additional test, we characterize the same troughs by their terminal velocities in this section. These are expected to be more agnostic to the intrinsic absorber shape and may therefore provide further insight.

\begin{figure*}
\gridline{
  \fig{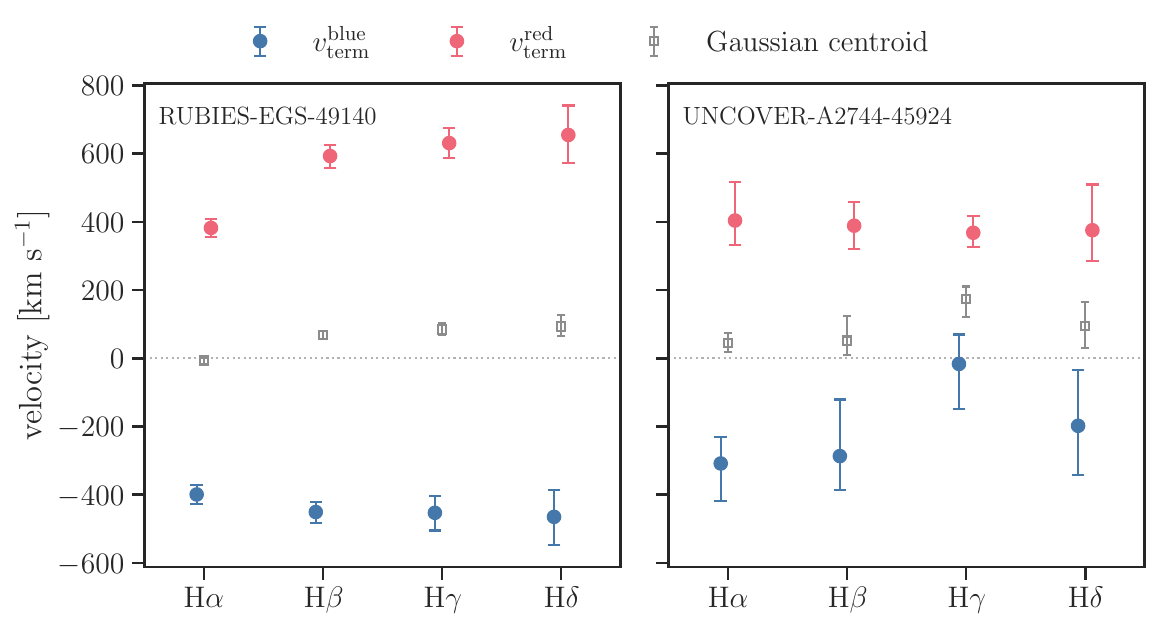}{0.95\textwidth}{}
}
\caption{Blue- and red-side terminal velocities (defined at the $95\%$ transmission level) of the Balmer absorbers in \rd\ and \ms\ as a function of transition (\ha--\hd). Error bars denote the 16th--84th percentiles, and the fitted Gaussian absorber centroids (open squares) are over-plotted for comparison. $\vtermb$ remains approximately constant across the Balmer series in both sources. $\vtermr$ is likewise consistent with being constant in \ms, but shows a mild systematic increase from \ha\ to \hd\ in \rd, consistent with the trend seen in the centroids.
}
\label{fig:vel_term}
\end{figure*}

The depth and resolution of our data resolve the absorption troughs well enough to measure both the blue- and red-side terminal velocities in all four Balmer transitions, up to \hd.  To our knowledge this is the first time that the full series has been presented, while previous measurements have largely focused on \ha\ and \hb. The results are shown in Figure~\ref{fig:vel_term}.

In both sources the blue terminal velocity is approximately constant along the series: $\vtermb \sim -500~\kms$ in \rd, and $\sim -300~\kms$ in \ms.
These values are comparable to those measured for the sample in \citet{Matthee2026}, for which \citet{Naidu2026:sn} report a median $\vtermb$ of $-443^{+43}_{-126}~\kms$.

The red terminal velocity is likewise consistent with being constant in \ms\ ($\sim +400~\kms$).
This behavior arises because the fitted absorber becomes broader toward the higher-order Balmer lines, while the absorption troughs simultaneously become shallower. Since the terminal velocities are defined at a fixed transmission level, these two effects largely offset one another, leaving them nearly constant.
The exception is \rd, where the red terminal velocity shows a mild increase from $+380~\kms$ at \ha\ to $+660~\kms$ at \hd, consistent with the trend seen in the absorption centroids.

\subsection{Helium Lines\label{sec:res:he}}

\subsubsection{Strong \hei\,$\lambda$7067, with a Broad Base\label{sec:res:hei}}

Broad \heib\ and \hei\,$\lambda$7067, both permitted triplet transitions, are clearly detected in \ms, with velocity offsets of $\sim 100~\kms$ and widths of $\sim2500~\kms$.
While \hei\,$\lambda$7067 falls outside the observable range for \rd, broad \heib\ is not detected, possibly because of insufficient S/N (Figure~\ref{fig:fit_oiii_neiii}).

The \hei\,$\lambda$7067/$\lambda$5877 ratio in \ms\ provides a density-sensitive diagnostic. Both lines are triplet transitions, but \heib\ is dominated by recombination, whereas \hei\,$\lambda$7067 has an additional important contribution from collisional excitation out of the metastable $2\,^3S$ level \citep{Berg2026}. Typical star-forming regions exhibit ratios of $\sim0.3$--$0.6$ (e.g., \citealt{Yanagisawa2026:atlas}). In \ms, we measure a \hei\,$\lambda$7067/$\lambda$5877 ratio of $1.10^{+0.14}_{-0.13}$ for the narrow component and $1.80^{+0.34}_{-0.29}$ for the broad component.
For comparison, a similarly high ratio of $\approx1.18$ has been reported for \fe\ based on a single-Lorentzian fit \citep{Torralba2026:fe}.
The enhanced \hei\,$\lambda$7067 emission therefore provides evidence for high densities. Additional line-ratio diagnostics are examined in Section~\ref{sec:res:ratios} and provide consistent support for the density diagnostic.

\subsubsection{Weak \heii}
\label{sec:res:heii}

\begin{figure*}
\gridline{
  \fig{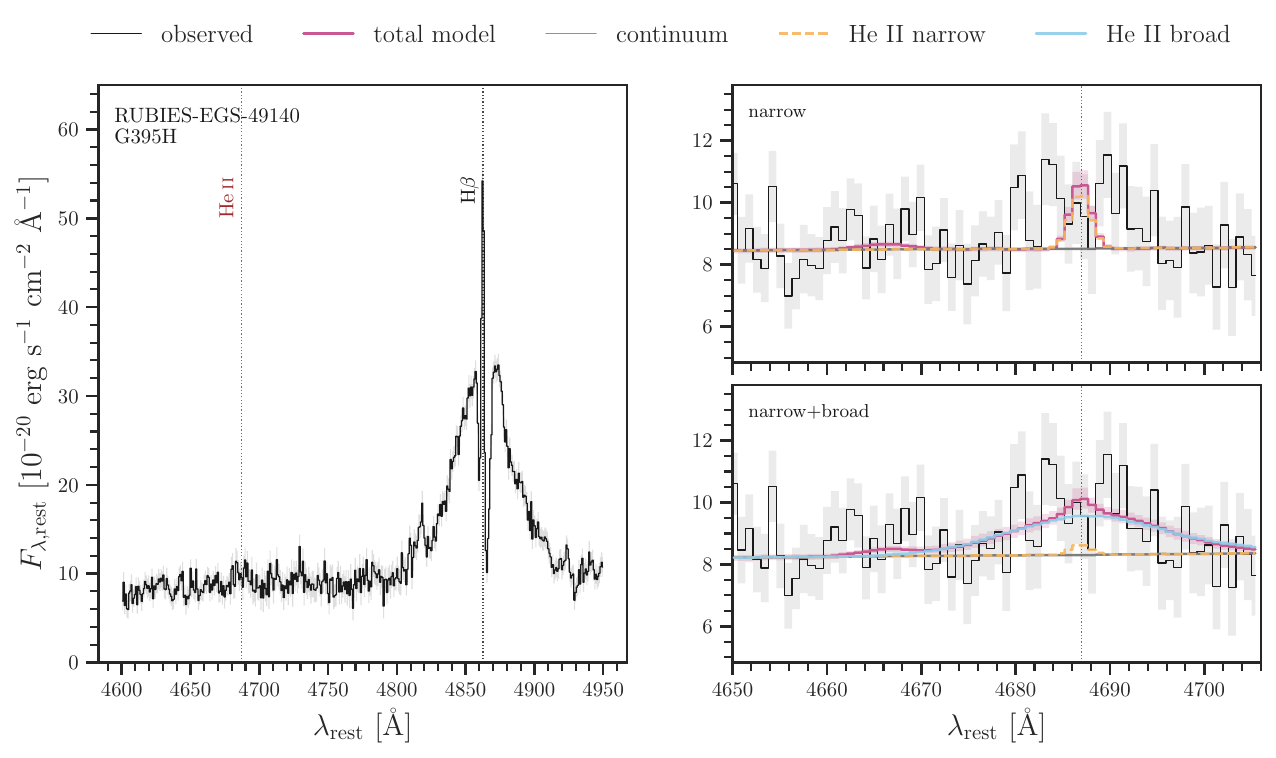}{0.45\textwidth}{}
  \fig{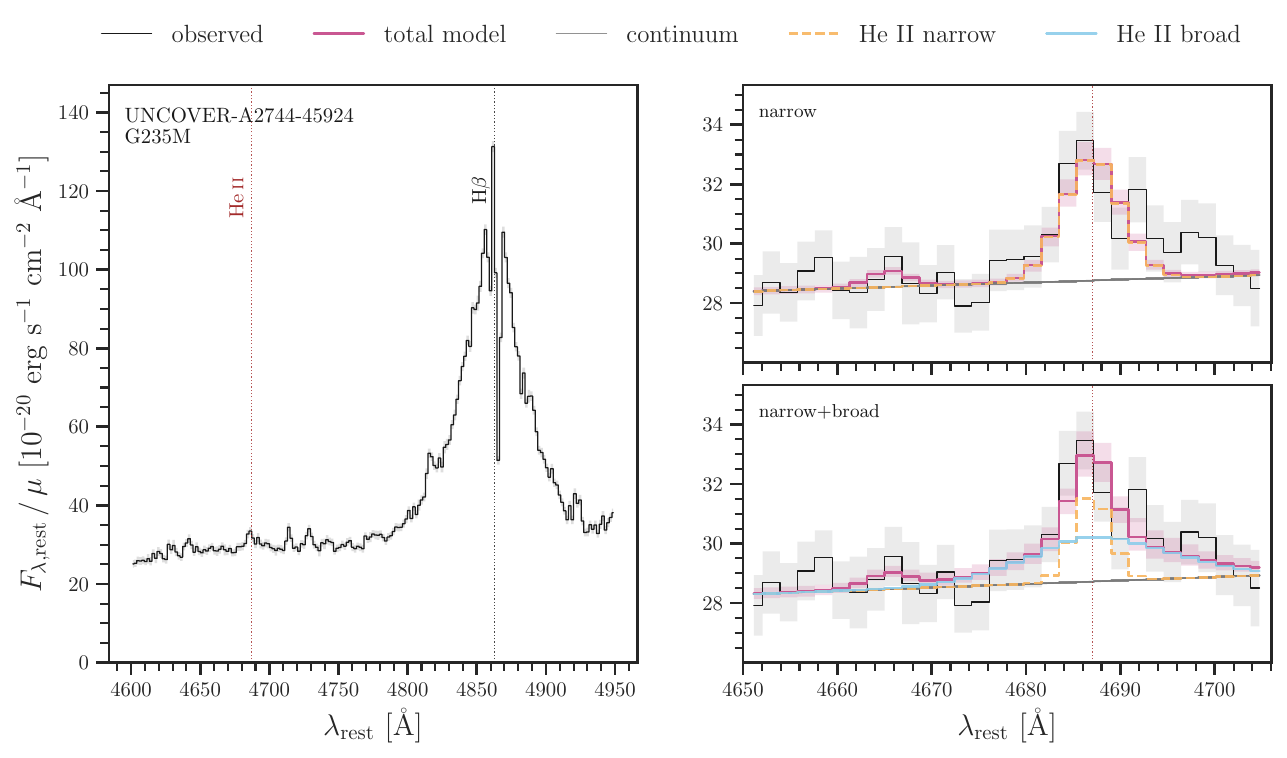}{0.45\textwidth}{}
}
\caption{Zoom-ins on the \heiir\ and \hb\ regions of \rd\ and \ms, illustrating the weak or undetected \heiir\ emission in both sources. For each source, we show the data together with two fits: a narrow Gaussian component only, and a narrow plus broad Gaussian component.
}
\label{fig:spec_heii_hb}
\end{figure*}

\heiir\ is very weak or undetected: it is not detected in \rd\ and is only marginally detected in \ms, despite the latter having exceptionally luminous \ha\ emission (Figure~\ref{fig:spec_heii_hb}). We estimate 95th-percentile upper limits of \heiir/\hb\ $<0.1$ for the narrow-line ratios, and $<0.007$ for the total flux ratios, for both sources. These limits confirm the previous finding based on medium-resolution and prism data \citep{Wang2026:qion}. The higher spectral resolution and exceptional S/N of the newly acquired \ms\ spectrum, together with the resolved \feii\ emission, further strengthen the conclusion that \heii\ is intrinsically weak.

One possibility raised by \citet{Wang2026:qion} is that \heii\ may be too broad to be readily detected. To account for this possibility, we additionally include a broad Gaussian component when fitting \heii. This yields 95th-percentile upper limits of \heiir/\hb\ $<0.03$ and $<0.01$ on the total flux ratios for \rd\ and \ms, respectively.

\begin{figure*}
\gridline{
  \fig{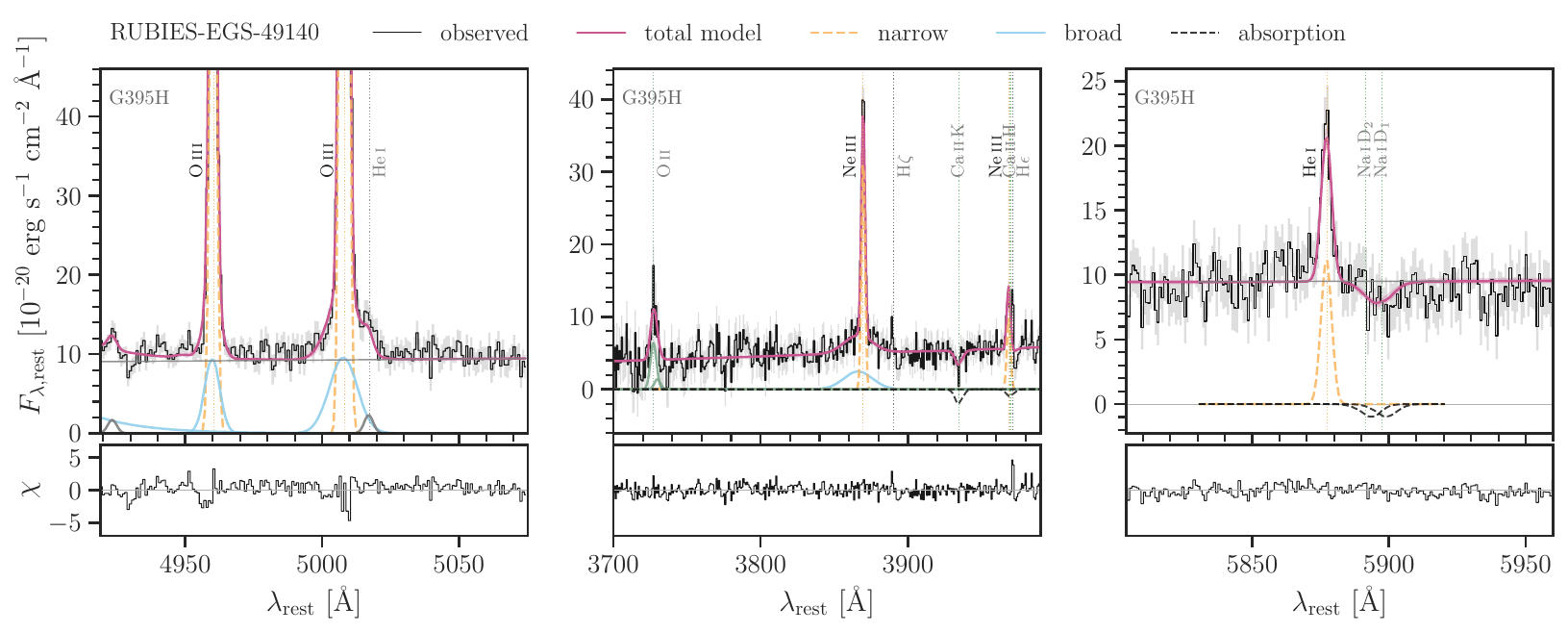}{0.95\textwidth}{(a)}
 }
\gridline{
  \fig{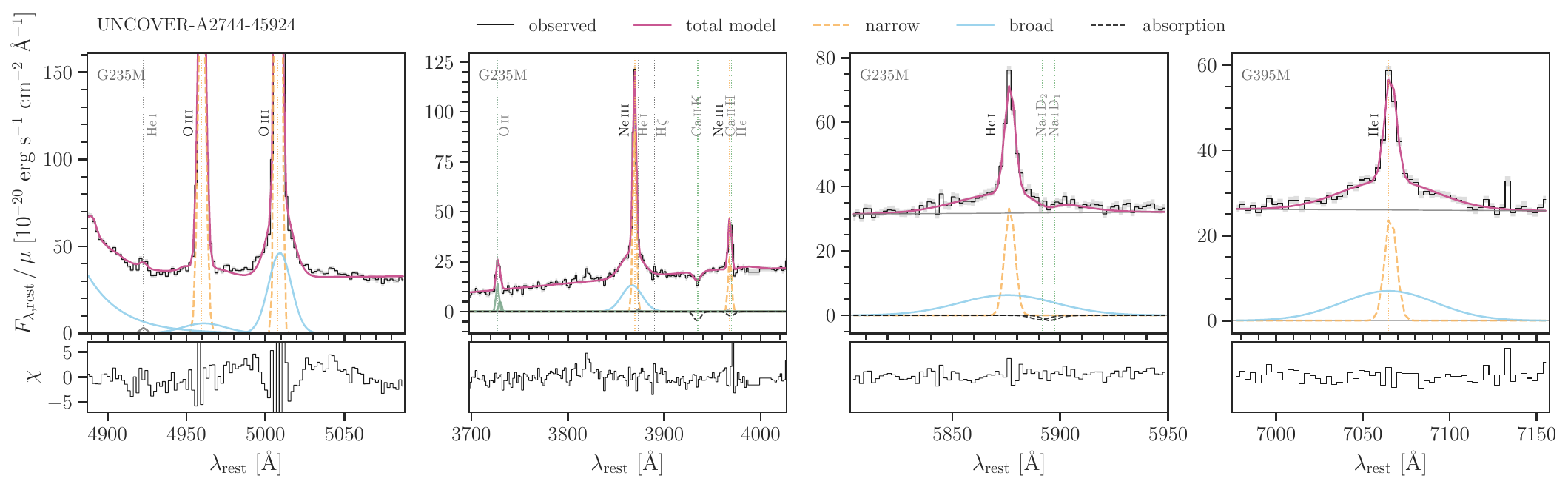}{0.95\textwidth}{(b)}
}
\caption{Model fits to the metal emission lines. (a) \rd: from left to right, the panels show the \oiii, \neiii, and \hei\ line complexes. (b) Same as (a), but for \ms. In both sources, \oiii\ and \neiii\ exhibit broad components, and Ca K absorption is detected.
}
\label{fig:fit_oiii_neiii}
\end{figure*}

\subsection{Metal Lines\label{sec:res:metal}}

\subsubsection{Broad \oiii, \neiii}

The metal-line model fits are shown in Figure~\ref{fig:fit_oiii_neiii}, with their fitted properties summarized alongside those of the Balmer lines in Figure~\ref{fig:components}.
We report, for the first time, broad \neiiiw\ components in two luminous LRDs. We also detect broad \oiiiauroral\ emission, consistent with recent results from medium-resolution spectroscopy \citep{Papovich2026}.

The narrow metal lines are near systemic as defined by narrow  \oiiidoubb. In \rd\ the narrow metal lines (\neiiiw, \oiiiauroral) sit at the systemic velocity, with $\fwhm \approx 200$--$300~\kms$. In \ms, they tend to be slightly blueshifted ($\sim-40~\kms$) relative to narrow \oiiidoubb, but consistent with systemic within $2\sigma$, with similar widths.

As mentioned in Section~\ref{sec:method:fit}, we also explore fitting the broad metal-line emission with exponential profiles instead of Gaussians, given that some of these broad components are reported here for the first time. The two models provide statistically indistinguishable fits (Appendix~\ref{sec:app:expnb}); we adopt Gaussian profiles as our fiducial model and quote the corresponding measurements below.

We detect a broad \oiiidoubb\ component consistent with the systemic velocity in both sources, with $\fwhm \approx 800$--$1500~\kms$. A broad \neiiiw\ component is also detected in both, blueshifted by $\sim 200~\kms$ and with $\fwhm \sim 1800~\kms$.

The two LRDs of this paper additionally require a broad \oiiiauroral\ component. In \ms, its presence is unambiguous owing to the exceptional S/N, with $\fwhm = 1887^{+83}_{-191}~\kms$ and $v = +29^{+123}_{-122}~\kms$. The deep G395H spectrum of \rd\ likewise favors a broad component ($\fwhm = 1689^{+195}_{-209}~\kms$, $v = +10^{+103}_{-83}~\kms$), the inclusion of which significantly improves the fit.

In both objects, the width of the broad \oiiiauroral\ matches that of the broad \neiiiw\ and is consistent within $1\sigma$ with that of the broad \oiiidoubb. The \oiiiauroral\ component is also unusually strong, as becomes evident when considering the line ratios in Section~\ref{sec:res:ratios}.

\subsubsection{Resolved \feii\ Emission}

The \feii\ multiplets are well modeled by a single kinematic system at the systemic velocity ($v \approx -1~\kms$ in \rd; $\approx-18~\kms$ in \ms) with $\fwhm \approx 404$ and $280~\kms$, respectively---intermediate between the narrow emission and the Balmer absorbers. A full analysis of the \feii\ spectrum is beyond the scope of this paper (cf.\ \citealt{DEugenio2025:fe, Lambrides2025:fevii}); we use it here primarily to obtain a clean continuum decomposition and for the line-identification test in Appendix~\ref{sec:app:fevii}.

With the emission-line decomposition established, we now proceed to construct component-by-component emission-line ratios to further investigate the ionization conditions in LRDs.

\subsection{Line Ratios\label{sec:res:ratios}}

\begin{figure*}
\gridline{
  \fig{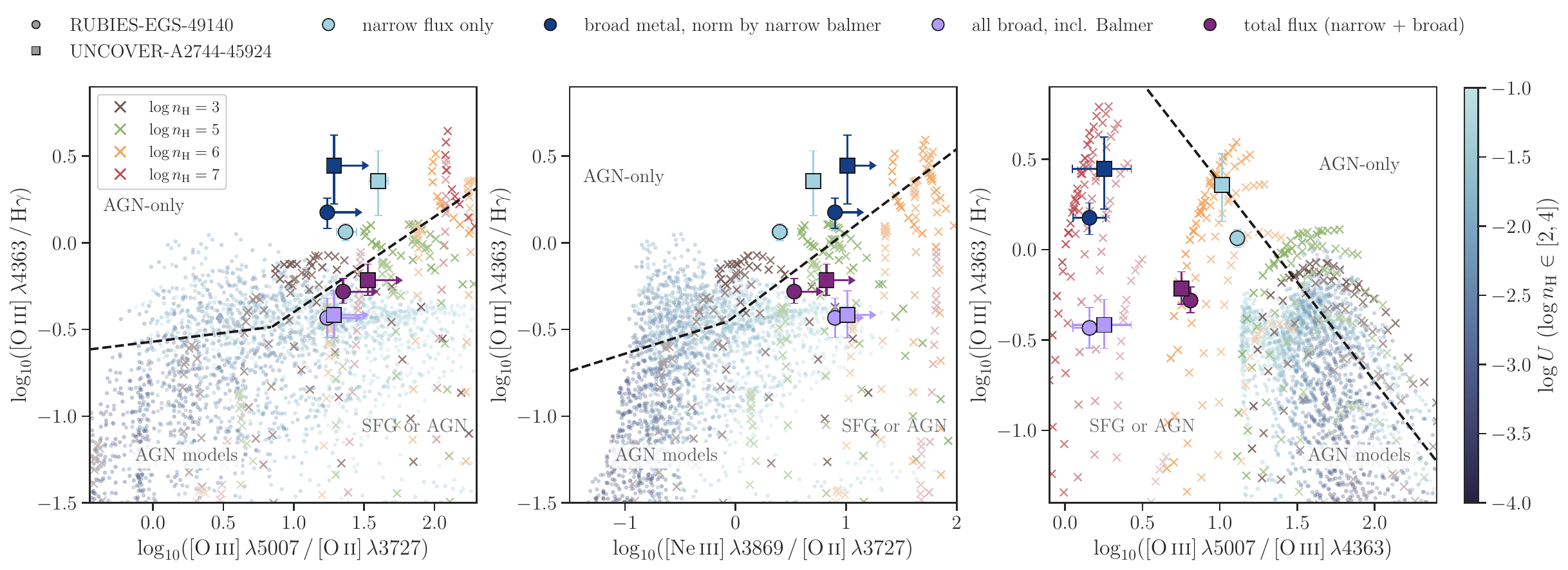}{\textwidth}{}
}
\caption{Line ratios diagnosing the photoionization conditions in \rd\ (circles) and \ms\ (squares). Colors indicate the flux components used to calculate the ratios: narrow flux only (light blue), broad metal-line fluxes normalized by the narrow Balmer flux (dark blue), broad fluxes (light purple), or total flux (dark purple). No broad component is detected in \oiin; ratios involving a broad or total \oiin\ flux are therefore 95$^{\rm th}$-percentile lower limits, shown as rightward arrows.
Background points are \cloudy\ photoionization models assuming standard AGN conditions \citep{Cleri2025}, color-coded by ionization parameter $\log U$; 
dashed curves mark the star-forming/AGN demarcation in each diagram \citep{Mazzolari2024}.
The narrow-line ratios of both LRDs fall outside the low density ($\log n_{\rm H} \in [2,4]$) model grids.
An independent set of NLR models (\citealt{Plat2026}; shown as crosses) extends to higher densities, spanning $\log n_{\rm H}=4$--$7$, and reaches the observed ratios only at $\log n_{\rm H} \gtrsim 5$.
Given the lack of very high-ionization lines, we interpret the failure of the standard AGN photoionization grids to reproduce the observed ratios primarily as a density effect.
}
\label{fig:ratios}
\end{figure*}

Emission-line ratios are a standard tool for identifying the ionizing source and constraining the physical conditions of the emitting gas. At high redshift, however, their diagnostic power is more limited because the location of a source on the classic optical diagnostic diagrams depends primarily on the ionization parameter ($\log U$) and gas-phase metallicity, rather than uniquely on the shape of the ionizing spectrum \citep{Cleri2025}. Even so, these diagnostics provide a useful reference by placing LRDs in the context of well-studied AGN and star-forming galaxy populations.

In the standard AGN framework \citep{Antonucci1993,Osterbrock2006,Netzer2015}, the optical emission lines originate in two physically distinct regions: an extended, lower-density narrow-line region (NLR), spanning hundreds to thousands of parsecs and producing the forbidden lines together with the narrow Balmer components, and a compact ($0.01$--$1$~pc), dense broad-line region (BLR), where strong forbidden lines are collisionally suppressed and only broad permitted (and semi-forbidden) lines survive. 

Motivated by this picture, we construct diagnostic ratios from the corresponding decomposed components. Figure~\ref{fig:ratios} shows three diagnostic diagrams \citep{Mazzolari2024} overlaid with AGN photoionization grids computed with \cloudy\ \citep{Cleri2025} spanning a broad range of ionization parameters, gas densities, and metallicities.

The narrow forbidden lines used in these diagrams—\oiin, \neiiiw, \oiii\,$\lambda$4364, and \oiiidoubb—are collisionally excited transitions with critical densities of $\necrit \sim 10^{3}$--$10^{7}\,{\rm cm^{-3}}$ ($\necrit \simeq 1.2\times10^{3}$ and $4.1\times10^{3}\,{\rm cm^{-3}}$ for \oii\,$\lambda$3730 and $\lambda$3727, $1.0\times10^{7}\,{\rm cm^{-3}}$ for \neiiiw, and $6.9\times10^{5}$ and $2.4\times10^{7}\,{\rm cm^{-3}}$ for \oiiidoubb\ and \oiii\,$\lambda$4364, computed with \pyneb\ \citep{Luridiana2015} at $T_e=10^{4}\,$K) and are thus suppressed by collisional de-excitation in denser gas. The broad Balmer wings, by contrast, are expected to arise in the much denser BLR, where forbidden-line emission is strongly suppressed.

It follows that ratios of forbidden-line flux to narrow Balmer-line flux provide a comparison of lines expected to originate in spatially similar regions. We accordingly consider two sets of ratios based on narrow emission: narrow fluxes only, and broad metal-line fluxes normalized by the flux of the narrow Balmer emission components. We caution, however, that this interpretation assumes an NLR/BLR-like separation between the forbidden- and permitted-line-emitting gas, which may not hold for LRDs. While these ratios are no longer directly comparable to the narrow-line model predictions, we include them in Figure~\ref{fig:ratios} for completeness.

In none of the diagnostic diagrams that use narrow Balmer emission for normalization do the AGN grids reach the observed ratios. Star-forming models can approach the data only at low ionization parameters ($\log U \approx -3$ to $-3.75$) and sub-solar metallicities, but these models fail to reproduce the observed \neiii/[O\,{\sc ii}] ratios. Thus, no standard low-density ($\log n_{\rm H} \in [2, 4]$) photoionization model simultaneously reproduces all of the observed diagnostics. This is broadly consistent with the offset of LRDs from star-forming \cloudy\ grids \citep{Sok2026}.
Pushing the models to higher densities ($\log n_{\rm H} > 4$) would bring the model tracks close to the observed data points. We confirm this by 
selecting points from the original low-density model grid of \citet{Cleri2025} and increasing $\log n_{\rm H}$ while holding the other model parameters fixed. 
An independent set of NLR models by \citet{Plat2026} supports this interpretation. Extending to $\log n_{\rm H}=3$--$7$, these \cloudy\ models reproduce the observed ratios only at $\log n_{\rm H} \gtrsim 5$. Following \citet{Feltre2016}, they adopt black-hole-mass- and Eddington-ratio-dependent accretion-disc SEDs from \texttt{relqso} \citep{Hagen2023}. The models shown here assume $\log(M_{\rm BH}/M_\odot)=7$ and $\log(\dot{M}/\dot{M}_{\rm Edd})=-0.5$.
Although the input spectra and model setups differ in detail, the two model sets \citep{Cleri2025, Plat2026} qualitatively agree that the observed line ratios favor high-density gas.
We note that these models are intended only to demonstrate high density as a possible explanation; a fully self-consistent exploration is beyond the scope of this work.

\section{Discussion\label{sec:dis}}

In this section, we first place the two sources studied here in the context of the broader LRD absorber population, summarizing what is distinctive about them as well as what they share with the broader population (Section~\ref{sec:dis:context}). We then turn to the constraints provided by the high-resolution, high-S/N spectra presented in this paper (Section~\ref{sec:dis:empirical}), and finally consider physical interpretations of the puzzling feature regarding the trends in absorber velocities across the Balmer series (Sections~\ref{sec:dis:geometry}--\ref{sec:dis:models}).

\subsection{A Brief Overview\label{sec:dis:context}}

Balmer absorption is now commonly reported in LRDs. However, \rd\ and \ms\ are distinguished by an unusual property: absorption at or redward of systemic velocity.
Near-systemic and redshifted absorbers are rare compared with the more common blueshifted population. 
Across different samples, Balmer-absorption incidence estimates for LRDs span roughly $\sim40 - 65$\% \citep{Matthee2026, Lin2026, Juodzbalis2026:census}.
However, near-systemic absorbers are hardest to detect due to possible emission infilling and limited resolution at $R\sim1000$, so their incidence is likely underestimated. The asymmetry of the incidence rates between blue- and red-shifted absorption is nonetheless robust according to \citet{Juodzbalis2026:census}.

Two additional systematic trends are worth discussing. First, the absorption depths can depart from the atomic optical-depth ratios. In particular, \hb\ absorption deeper than \ha\ has been reported in \fe\ \citep{Torralba2026:fe}, in \qso\ \citep{DEugenio2026:qso1}, and in the sample-averaged optical depths of \citet{Matthee2026}. Interpreting these depths has so far been limited by degeneracies in the absorption models. \citet{Matthee2026}, accordingly, do not interpret their fitted $\tau_0$. \citet{Chen2026:abcd}, however, fit \ha--\hg\ with a partial-covering screen having a single covering fraction and atomically tied optical depths, finding that five of their six sources with absorption are consistent with this model. The exception is \rd.

Second, the velocity progression along the Balmer series has been reported for \qso\ \citep{Furtak2024} in \citet{DEugenio2026:qso1}, \fe\ in \citet{Torralba2026:fe}, for RUBIES-EGS-55604 \citep{Wang2024:ub} and RUBIES-EGS-42046 in \citet{Matthee2026}, and in a small fraction of the sample studied in \citet{Yanagisawa2026:atlas}.
\citet{Naidu2026:sn} also briefly note this behavior, with absorber velocity offsets tending to decrease down the Balmer series, such that higher-order lines are absorbed closer to systemic. This progression generally runs from blueshifted toward systemic velocities. In \rd\ and \ms, however, the same progression extends from systemic to redshifted velocities. Whether this progression represents a systematic property of the population likely requires a larger sample to establish.

Finally, the fact that the broad \oiiidoubb\ components sit near the systematic is also worth highlighting. We discuss their implications further below.

\subsection{Empirical Constraints from the Spectra\label{sec:dis:empirical}}

\subsubsection{The Origin of Broad \oiiidoubb}

In the local LRD sample of \citet{Lin2026:desi}, broad \oiiidoubb\ is detected in 21 of 27 objects, with median $\fwhm=265~\kms$ (range $173-592~\kms$). The components in \rd\ and \ms\ are broader, with $\fwhm\approx800~\kms$ and $\approx1500~\kms$, respectively. It remains unclear whether such components are common at high redshift. Broad \oiiidoubb\ has been reported in individual sources \citep[e.g.,][]{Juodzbalis2024:rs}, but no incidence rate has yet been established, since detecting a weak component of this width requires deep, medium- to high-resolution spectroscopy. The presence of a broad component carrying $\sim 10-20$\% of the total \oiiidoubb\ flux nevertheless raises the question of its physical origin.

The origin of \oiiidoubb\ emission in LRDs has been attributed to the host galaxy \citep[e.g.,][]{Sun2026}, consistent with their rest-UV morphologies often being more extended than the rest-optical emission \citep[e.g.,][]{Baggen2024, Killi2024, Chen2025, Wang2025:brd, Ando2026, Cloonan2026, Zhang2026:uv} and/or with the presence of a blue companion \citep[e.g.,][]{Baggen2026, Golubchik2026}.
Broad \oiiidoubb\ emission, in contrast, is more commonly associated with a non-stellar ionizing source \citep[e.g.,][]{Heckman1981, Greene2005, Mullaney2013, Zakamska2014, Woo2016} and could therefore indicate a contribution from the central engine. As a forbidden transition, its broadening traces bulk motions in lower-density ionized gas. AGN-driven outflows easily reach $\gtrsim 1000~\kms$ \citep[e.g.,][]{ForsterSchreiber2019}, consistent with the FWHMs measured for the two LRDs studied in this paper.

A complication is that broad \oiiidoubb\ is not unique to AGN. 
It is found in about half of the Ly$\alpha$ emitters at $z\approx2$ with \oiii\ detections \citep{Matthee2021}, and appears ubiquitously in the Lyman-continuum-emitting Green Pea galaxies at $z\sim0.3$ \citep{Amorin2024}. In both samples, the broad components are interpreted as ionized outflows driven by massive stars and supernova feedback. \citet{Berg2026:wr} more recently reported broad Balmer and \oiiidoubb\ components with $\fwhm\approx700-3000~\kms$ in RXCJ2248, a lensed, compact ($R_{\rm eff}\approx20$~pc), metal-poor star-forming region at $z=6.1$, where the broad emission is powered by Wolf-Rayet nitrogen stars. Winds from very massive stars can potentially produce similar widths, as invoked to explain the broad N\,\textsc{iv}] emission ($\fwhm\approx1670~\kms$) in GN-z11 \citep{Chen2026}.
We caution, however, that this resemblance does not by itself establish a host-galaxy or stellar origin. We only note that the widths measured for the two LRDs of this paper fall within a regime that can, in principle, be produced by sufficiently dense and massive star clusters.

\subsubsection{Anomalous Line Ratios\label{sec:dis:linratios}}

We interpret the failure of the AGN photoionization grids to reproduce the line ratios of \rd\ and \ms\ primarily as a density effect. The forbidden lines remain optically thin and are collisionally excited at rates set by the electron temperature, but above the critical density, the upper level is more likely to be de-excited by a subsequent electron collision before it can radiate. Because \oiin\ has the lowest critical density and \oiiiauroral\ the highest, increasing the density suppresses these lines in the order required by the observed ratios: \neiiiw/\oiin\ and \oiiiauroral/\oiiidoubb\ both increase as the density rises. This suggests that the forbidden lines arise in gas substantially denser than a classical NLR, but still below the densities inferred for the Balmer-absorbing gas.

Such high-density interpretation is additionally supported by the \hei\ line ratios, consistent with the picture of dense gas proposed by \citet{Torralba2026:fe}, and is strengthened by our resolved narrow- and broad-line \hei\ decompositions, rather than single-profile fits.
A detailed treatment of \hei\ radiative transfer is nevertheless non-trivial: \citet{Torralba2026:fe} found that their simple \cloudy\ models fail to reproduce the observed line ratios. A dedicated model grid exploring the relevant density and radiative-transfer effects will therefore be needed to constrain the physical conditions, which we leave to future work.

\subsubsection{Transition-dependent Absorber Velocities and Depths}
\label{sec:dis:tau}

We detect and resolve Balmer absorption in all four transitions. Notably, the absorber centroids shift from near the systemic velocity in \ha\ to increasingly redshifted velocities in the higher-order lines in \rd\ (Section~\ref{sec:res:balmer}). This is unexpected under the common assumption of a single homogeneous absorbing gas slab (e.g., \citealt{Ji2025, Naidu2025, Wang2026:qion}), which would imprint the same velocity structure on every Balmer transition.

An independent constraint comes from \caiik, a resonance transition of singly ionized calcium that traces neutral and weakly ionized gas, which we detect in absorption in both sources. In \rd, the \caii\,K trough is centered near the systemic velocity, with $v=-46^{+39}_{-41}~\kms$, consistent within uncertainties with the \ha\ absorption but offset from the redshifted higher-order Balmer troughs. In \ms, it is likewise consistent with being near systemic ($v=-58^{+134}_{-97}~\kms$).
The finding that the gas traced by \caii\ K follows the lowest-order Balmer absorption, rather than the redshifted \hb--\hd\ troughs, suggests that the higher-order Balmer lines probe an additional absorbing component that is not traced by the calcium phase, or a smooth gas medium with a velocity gradient with the \caii\ K and \ha\ absorptions emerging at different depths from \hb--\hd. We further discuss the possible physical origins in Section~\ref{sec:dis:models}.

The absorption depths show a second, independent departure from the single-screen expectation. All Balmer transitions arise from absorption out of the $n=2$ level, such that, under the assumption of a common covering fraction, their relative optical depths are fixed by atomic physics, scaling as $f\lambda$ (Section~\ref{sec:intro}): $\tau_{\ham}/\tau_{\hbm} = (f\lambda)_{\ham}/(f\lambda)_{\hbm} \approx 7.25$ and $\tau_{\hbm}/\tau_{\hgm} \approx 2.99$.
The measured troughs are too similar in strength, with the absorbed EWs declining only mildly along the series: from $9.4$ to $5.5$~\AA\ between \ha\ and \hd\ in \rd, and from $8.9$ to $2.6$~\AA\ in \ms\ (Table~\ref{tab:balmer}). In particular, EW$/\lambda$ appears weaker in \ha\ ($-1.42^{+0.14}_{-0.24}\times10^{-3}$) than in \hb\ ($-1.63^{+0.12}_{-0.12}\times10^{-3}$) or \hg\ ($-1.52^{+0.21}_{-0.21}\times10^{-3}$) in \rd. Saturation alone cannot account for this, since EW$/\lambda$ of an attenuating screen grows monotonically with $\lambda f n_2/b$ and therefore leaves \ha\ the deepest trough at any $n=2$ column density $n_2$. \citet{Chen2026:abcd} noted this tension for \rd\ among their sample of six LRDs with absorption detected with medium-resolution data; our resolved decomposition confirms it, and extends the measurement to \hg\ and \hd\ and to \ms.

Rather than indicating a breakdown of atomic physics, the pattern can be interpreted in two ways. The depths may encode a genuinely structured absorber, in which the observed line depths are modified by transition-dependent covering fractions and/or multiple kinematic components (e.g., \citealt{DEugenio2025:fe}). Alternatively, the absorbing gas may not act as a pure attenuator: the optically thick absorbing medium plausibly emits light in its own right with a non-negligible source function, which preferentially fills in the low-order Balmer absorption due to radiative trapping effects (H. Liu et al., in preparation). A similar principle underlies the shallow H$\alpha$ absorption seen in A-type main-sequence stars and could explain the observed absorption depth patterns in LRDs without invoking complex gas geometry. Viewed from another perspective, the source function may provide a physical basis for the phenomenologically motivated transition dependence of the covering factor, whose underlying origin remains uncertain.

\subsection{Break Strength as the Potential Fundamental Correlate\label{sec:dis:geometry}}

\rd\ and \ms\ are similar in the optical: both are among the reddest (in terms of the ``slope'' between the Balmer limit and $\sim 5500$~\AA) and most luminous LRDs, both have strong Balmer breaks, and both, as shown in this work, have near systemic Balmer absorbers together with broad metal line components. It is thus natural to ask whether this points to certain common geometric origin.

Wind models are built for the P-Cygni class: \citet{Madau2026} and \citet{Naidu2026:sn} model blueshifted troughs from a slow outflow. Systemic absorbers appear in these frameworks only as outliers, attributed to turbulence or rotational broadening of a slow wind; i.e., yet to be treated in a physically consistent way within the same framework.

Viewing-angle models offer a partial answer. \citet{Matthee2026} find that redder LRDs more commonly exhibit systemic Balmer absorption, and propose a biconical geometry---structurally similar to the cocoon model of \citet{Sneppen2026}---in which bluer LRDs are viewed through the outflow cone (yielding blueshifted absorbers), while redder sources are viewed through the thicker equatorial gas (yielding systemic absorption). In this case, the most luminous \ha\ emitters and the strongest Balmer breaks correspond to envelopes with the highest column density, which are also the most likely to show absorption.
However, viewing angle alone does not explain why the systemic absorbers are also among the most luminous sources. \citet{Matthee2026} additionally discuss an alternative evolutionary phase in which the column density of the envelope, and hence whether the absorption is systemic, changes over the source lifetime.

A complication is that the apparent red--luminous connection may partly be selection. The v-shape SED selection may bias against red, low-luminosity sources. For instance, \qso\ \citep{Furtak2024} is intrinsically much fainter and discovered with the aid of strong lensing. As noted in Section~\ref{sec:dis:context}, it shows similar behavior as the two sources of this paper: its \ha\ and \hb\ absorbers are both close to systemic, with \ha\ slightly blueshifted ($-40\pm10~\kms$) and \hb\ slightly redshifted ($+50\pm20~\kms$), and the relative absorption strengths also depart from the trend expected from atomic physics \citep{DEugenio2026:qso1}.
However, selection introduces another puzzle. The complementary region of parameter space---sources as luminous as \rd\ and \ms\ but with shallower optical slopes---should be easy to detect. One such example may be GS-3073 \citep{Ubler2023, Ji2024}, which \citet{Matthee2026} identify as the luminous, blue exception in their sample. The apparent absence of such luminous, ``bluer'' LRDs therefore suggests that the observed correlation is unlikely to be purely an artifact.

Importantly, \qso\ is far less luminous than \rd\ and \ms, yet shares with them both a strong Balmer break and near-systemic absorbers, with more blueshifted absorption in \ha\ and more redshifted absorption in \hb\ \citep{DEugenio2026:qso1}.
Therefore, the strength of Balmer break may be a more fundamental correlate.

We stress, however, that the above interpretation is based on only a handful of objects. The small sample of LRDs with well-characterized absorber velocities makes it difficult to disentangle break strength from correlated properties, and absorbers are particularly challenging to measure in fainter sources. The local LRD analog, the ``Egg'' \citep{Lin2026}, similarly shows blueshifted \ha\ and redshifted \hb\ absorption, but has a much weaker Balmer break, although the latter may partly be a manifestation of local LRDs being possibly more susceptible to host-galaxy contamination.

\subsection{Possible Physical Origins\label{sec:dis:models}}

The picture considered in \citet{Matthee2026} is structurally similar to the cocoon model of \citet{Sneppen2026}, but the latter is implemented with \sirocco\ radiative-transfer models, with polar outflows and equatorial inflows. Different viewing angles can therefore produce P-Cygni, inverse P-Cygni, and near-systemic absorption profiles.
Intriguingly, however, the model sequence used to span the observed range of break strengths predicts the opposite trend to that inferred by \citet{Matthee2026}: the strongest breaks correspond to the most blueshifted absorption.

It is also important to note that, as discussed by \citet{Sneppen2026}, an inflowing component is required to reproduce the observed line profiles, but their model does not provide a physical explanation for its coexistence with the outflow. This motivates the question we focus on in this discussion: what physical mechanism could produce the inflows?

\subsubsection{T Tauri Stars as a Precedent for Velocity Shifts in Absorption\label{sec:dis:ttauri}}

We first ask whether transition-dependent absorber velocities, changing from an outflow signature in \ha\ to an inflow signature in the higher-order Balmer lines, have been observed in other systems.
A precedent exists in T Tauri stars (e.g., \citealt{Appenzeller1988, Edwards1994, Alencar2000}). In classical T Tauri stars (CTTSs), the Balmer absorption is understood in the context of magnetospheric accretion (see \citealt{Bertout1989} for a review). The disc is truncated at a few stellar radii by the stellar magnetic field, and material is lifted from the disc and channeled along field lines, accelerating toward the stellar surface and approaching free-fall velocities before the accretion shock \citep{Hartmann1994, Muzerolle2001}.

Here, the absorbing gas is the infalling accretion column, while the disc and shocked surface provide the background continuum \citep{Edwards1994, Hartmann1994}. Redshifted troughs appear when a funnel stream is projected against the hot accretion spot, while against the cooler photosphere the funnel flow does not produce sufficient line contrast to form a trough. Consequently, the strength of the redshifted absorption depends sensitively on inclination and on the extent of the magnetospheric accretion zone \citep{Hartmann1994, Alencar2000}. Blueshifted absorption, when present, instead requires a wind launched exterior to the magnetosphere---an accretion-powered stellar wind or disc wind---and cannot be produced by the funnel flow \citep{Muzerolle2001, Kwan2007}.
The metastable \hei\,$\lambda$10830 line is also a sensitive tracer of both flows \citep{Edwards2003, Kwan2007}, with both blueshifted and redshifted absorption reported. While it has so far been accessible in only a handful of LRDs at $z\lesssim3.7$, P-Cygni absorption features have been observed (e.g., \citealt{Juodzbalis2024:rs, Wang2025:brd, Kokorev2026:glimpse, Lin2026:egg}).

Quantitatively, the LRD absorber velocities are of the same order of magnitude as those for CTTSs. In magnetospheric infall models, gas free falls from the disk truncation radius and reaches $\sim300~\kms$ by the time it shocks the stellar surface \citep{Hartmann1994, Muzerolle2001, Hartmann2016}; in the hottest, highest velocity tracers (e.g., \ion{C}{4}), formed closer to the shock, infall signatures have been reported up to $\sim400~\kms$ \citep{Hartmann2016}. The Balmer absorption $\vabs$ measured for the two LRDs reach $\sim100~\kms$, slightly below but still within the same order of magnitude as these characteristic infall velocities. The absorber widths ($\fwhm\approx200$--$500~\kms$) are likewise comparable to the CTTS free-fall velocities.

Besides redshifted absorption, diverse \ha\ emission-line morphologies are also observed in T Tauri stars, generally understood as arising from the interplay between accretion and outflow. Models such as \citet{Kurosawa2006,Lima2010} explicitly include both a magnetospheric accretion flow and a disc wind, together with the central continuum source and accretion disc. The disc wind is a biconical outflow launched outside the magnetosphere, while the relative contributions of the inflow and outflow to the \ha\ profile depend on the accretion rate, wind structure, and viewing angle.
Thus, a coupled inflow--outflow geometry can naturally produce diverse \ha\ profiles.

Stellar analogies for LRDs are a relatively recent development \citep[e.g.,][]{Chisholm2026:gc, Martins2026, Nandal2026, Naidu2026:sn}. 
The comparison presented here is not meant to be taken literally; see \citealt{Takasao2026} for a theoretical take on a similar analogy.
Rather, the relevant insight is that different Balmer transitions may preferentially sample different regions of a structured accretion flow and wind. A transition-dependent change in absorption velocity, from blueshifted to redshifted across the Balmer series, can therefore be accommodated within a self-consistent physical framework, albeit at much higher luminosity scale. We consider this possibility further below.

\subsubsection{A Rotating Disc Wind\label{sec:dis:rotation}}

A related scenario was recently invoked by \citet{Torralba2026:panbh} to explain the Balmer absorber in PAN-BH*-1, in which \ha\ absorption trough extends from $-520$ to $+267~\kms$ relative to systemic. Absorption spanning both sides of systemic cannot be produced by a simple radial flow. They instead find that a thick wind launched from a rotating, geometrically thick disc can produce such a profile in a dynamically steady configuration (i.e., no absorption variability on the dynamical timescale). The wind inherits an azimuthal velocity component ($\mathbf{v}_\phi$) from the rotating disc, so that at high inclination the projected velocity spans both blueshifted and redshifted values.
We further note that a rotating disc-wind scenario raises another possibility that the central trough is not absorption at all, but rather the minimum between two rotationally split emission peaks. \citet{Knigge1995} showed that a rotating accretion-disc wind can produce a narrow central core flanked by two peaks: their pure-rotation reference case, although idealized, produces a profile of this form, while their model with slow initial launching followed by gradual acceleration (their Figure~9) more closely resembles the LRD line shapes. A rotation-dominated disc wind could therefore mimic a systemic absorption trough.
Such a profile need not be the same across the Balmer series; as shown in e.g., \citet{Knigge1995}, the line profiles depend strongly on the density structure of the wind.

However, in the optically thick, quasi-spherical scattering-envelope scenario often invoked for LRDs, the line-emitting region is often expected to extend over a near-$4\pi$ solid angle, as argued in e.g., \citet{Yanagisawa2026:atlas, Kageura2026, Yan2026}, and across one to two orders of magnitude in radius. The emitting gas would therefore sample a much broader range of projected rotational velocities than a thin, radially compact BLR. In addition, polar streamlines in a Keplerian velocity field rotate more slowly, and repeated scattering can further redistribute and reprocess the photon directions. These effects would tend to wash out the sharp double-peak signature expected from an edge-on, geometrically thin rotating disc, although an analogous scenario involving a rapidly rotating massive star losing its outer layers or driving a wind may be less susceptible to this effect.
Notably, the rotationally supported models of \citet{Sneppen2026} do not predict prominent double-peaked profiles.

We caution, however, that the scattering-envelope interpretation, while well motivated, remains unproven. Notably, recent spectropolarimetry of a local LRD finds a $\sim 50^\circ$ offset between the continuum and broad-line polarization angles, indicating broken axial symmetry \citep{DEugenio2026:pol}. A rotation-dominated disc wind, in which the observed narrow emission-like core and absorption feature arise from the same rotating structure, therefore remains an intriguing possibility.

\subsubsection{A Failed Wind\label{sec:dis:failedwind}}

\begin{figure*}
\gridline{
  \fig{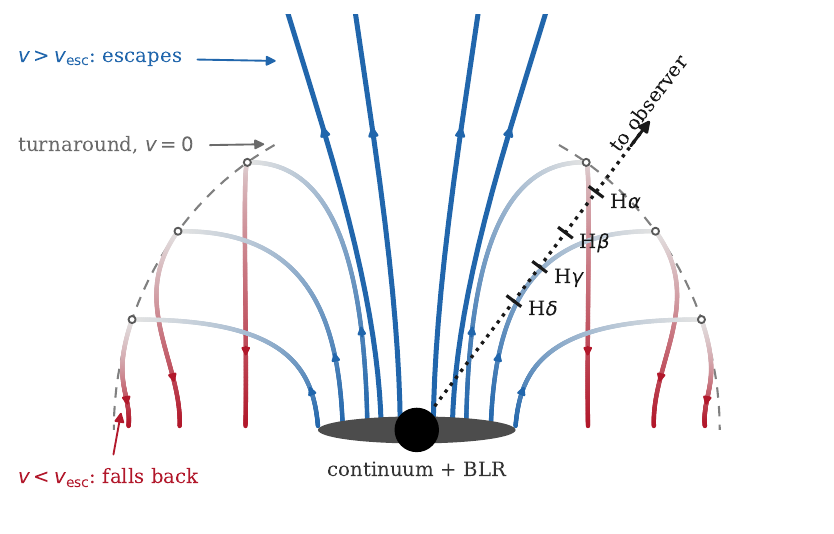}{0.65\textwidth}{}
}
\caption{Schematic illustration of a failed wind.
  Material launched along the pole above the escape speed (dark blue) escapes, while material launched at larger angles with $v<v_{\rm esc}$ decelerates, reaches $v=0$ on a turnaround surface (dashed), and falls back (red); color saturation tracks $|v|$.
  A sightline through the failed component crosses slow gas at large radius and faster gas deeper in.
  \ha, which forms in the outermost, fully decelerated layer, therefore absorbs near the systemic velocity, while the higher-order transitions probe deeper, faster-moving gas.
  The sign of the shift depends on whether the sightline samples the outflowing or the infalling branch; the redward progression measured in this paper (Section~\ref{sec:res:balmer}) requires the infalling side to dominate.
}
\label{fig:wind}
\end{figure*}

A conceptually similar picture to the T Tauri star analogy above is a decelerating, failed wind. For instance, radiation-driven disc-wind simulations show that gas can be lifted from the disc by the local UV continuum but becomes overionized by the central X-ray source before it can accelerate to the escape speed (e.g., \citealt{Murray1995, Proga2000, Risaliti2010, Mizumoto2021}). However, this situation requires higher energy photons than are seen in LRDs (e.g., \citealt{Wang2026:qion, Zucchi2026}). Another possibility is that the winds are launched by radiation pressure on electrons. This is analogous to Eddington-limited winds from massive stars, which can fail at sufficiently high mass-loss rates \citep{owocki_1997}. Whatever the case, the relevance here is that a failed wind would naturally provides a mechanism for the monotonic redward progression of the absorber velocity up the Balmer series.

By definition, a failed wind consists of streamlines launched with an outward radial velocity insufficient to escape: the gas decelerates, reaches the apex of its trajectory at $v=0$, and subsequently falls back (Figure~\ref{fig:wind}).
We note that it is not the entire wind that fails. Over most latitudes the gas is accelerated past the escape speed and leaves the system as an outflow; only at the lowest latitudes do the streamlines fail to reach $v_{\rm esc}$ and fall back. Because the escaping and infalling regions are contiguous in launch angle, there is necessarily a layer between them in which the outward velocity passes through zero. It is this turnaround layer, rather than the wind as a whole, that a near-systemic absorber requires the sightline to intersect.

In such a flow the absorbing gas is velocity-stratified with depth: \ha\ forms in the outermost, fully decelerated layer and absorbs near the systemic velocity, while the higher-order transitions probe deeper layers. 
The reason different transitions sample different depths is set by atomic physics. As noted earlier, the line-center optical depth of a Balmer transition scales as $\tau_0 \propto n_2\,f\,\lambda$. Since $f$ drops steeply along the series, a higher-order transition requires a correspondingly larger $n=2$ column to reach the same optical depth, and therefore necessarily thermalizes at deeper radii within the absorbing layer.
Given the velocity profile and the $n=2$ density profile, this stratification maps onto a per-transition velocity offset, potentially reproducing the observed trend.

The above scenario also connects the systemic absorbers to the broader P-Cygni population. Both would arise from the same physical structure viewed at different angles. Observing $v \approx 0$ at the trough requires a sightline that intersects the turnaround, so the relative rarity of systemic absorbers becomes a matter of viewing angle, which can be a sample-level prediction potentially testable against the observed distribution of absorber velocities.

The failed-wind model has recently been applied to LRDs (N. Kaaz et al., in preparation).
We emphasize, however, that current failed-wind models have not yet been tested against the specific line profiles observed here, and establishing whether they can quantitatively reproduce the data is beyond the scope of this work. Here, we simply present the failed-wind interpretation as a possible physical picture.

\subsubsection{Potential Discriminants}

A remaining question is how the various scenarios can be distinguished.
Although their underlying physics differs, several produce qualitatively similar phenomenology. Given that the data presented here are perhaps close to the best that can be hoped for, detailed modeling may be the most effective way forward, both for testing whether individual scenarios can reproduce the observed line profiles and for making population-level predictions that can be compared with the observed incidence and trends of absorbers. The former requires solving the radiative transfer and kinematics together, with a single model reproducing \ha--\hd\ simultaneously. Further constraints can come from other transitions, such as He and O\,\textsc{i}, which probe different excitation and ionization states, as well as from emerging absorption diagnostics, such as the Ca\,\textsc{ii} triplet (J. Greene et al., in preparation) and molecular absorption \citep{Wang2026:h2o}.
Population-level predictions may be equally constraining: each geometry implies a distribution over viewing angle, and therefore predicts the relative frequency of different line profiles, including near-systemic absorption versus P-Cygni profiles.

\section{Conclusions\label{sec:concl}}

This paper presents deep, high-resolution (G395H) JWST/NIRSpec spectroscopy of \rd\ ($z=6.68$) and high signal-to-noise medium-resolution spectroscopy of \ms\ ($z=4.46$), two of the most luminous LRDs known with strong Balmer breaks, covering all four Balmer lines \ha--\hd. The main findings are as follows.

First, we resolve Balmer absorption in all four transitions in both sources. The narrow and broad emission components lie close to the systemic velocity, and neither source shows the blueshifted P-Cygni profiles commonly seen in the LRD population. 

Second, both the velocities and the depths of the absorbers vary systematically along the Balmer series. The absorber centroid becomes more redshifted with increasing transition order, particularly in \rd. The relative absorption strengths, meanwhile, do not follow the naive expectations for a single absorbing screen set by atomic physics. The absorbed EWs decline far less than the atomic optical-depth ratios permit.

Third, \caiik\ is detected in absorption in both sources and is near systemic, tracking the \ha\ trough rather than the redshifted \hb--\hd\ troughs. Na\,D absorption is also detected in both sources, albeit with larger uncertainties on the kinematics.

Fourth, we report broad \neiiiw\ for the first time, together with broad \oiiiauroral\ and broad \oiiidoubb\ in both sources and broad \hei\ in \ms.
Intriguingly, the broad \oiiidoubb\ components are consistent with the systemic velocity.
Moreover, these detections enable a component-by-component examination of the standard line-ratio diagnostics. Standard AGN photoionization grids do not reproduce the observed narrow-line ratios at any ionization parameter, whereas increasing the gas density brings the model tracks closer to the data. The \hei\,$\lambda$7067/$\lambda$5877 ratio in \ms\ is also anomalously high, providing additional evidence for high densities.

Fifth, together with \qso, which is far less luminous yet shares both properties, the high-$z$ LRDs known to host near-systemic absorbers all have strong Balmer breaks. This suggests break strength, rather than luminosity or color, as the more fundamental correlate.

Finally, we postulate physical interpretations for the two trends along the Balmer series. 
While alternative scenarios stay open, we suggest two new interpretations. Regarding the relative absorption strengths, the self-emission of the optically thick absorbing medium could rescale each transition differently, which suppresses \ha\ absorption relative to \hb\ without requiring a transition-dependent covering fraction. The redward progression of the absorber velocity, in turn, can arise in a stratified, decelerating failed wind, in which the higher-order transitions probe deeper, faster-moving layers.

In summary, the high-resolution and high-S/N spectroscopy presented in this paper provides a detailed view of the kinematics, physical conditions, and structure of the line-emitting and absorbing gas in LRDs. Detailed models that connect the observed line profiles to the underlying gas dynamics and physical conditions will be a critical next step toward understanding this population.

\begin{acknowledgments}

B.W. thanks Dan Coe, the NIRSpec reviewer for PID 8047, for invaluable assistance in designing the MSA observation; Andrea Weibel for help with the photometry catalog; and Olivia Curtis, Francesco D'Eugenio and Lizhong Zhang for insightful discussions.
B.W. acknowledges support provided by NASA through Hubble Fellowship grant HST-HF2-51592.001 awarded by the Space Telescope Science Institute (STScI), which is operated by the Association of Universities for Research in Astronomy, In., for NASA, under the contract NAS 5-26555.
A.Z. acknowledges support by the Israel Science Foundation Grant No. 864/23.

Support for Program No. JWST-GO-08047.001-A was provided through a grant from the STScI under NASA contract NAS5-03127.
This work is based on observations made with the NASA/ESA/CSA James Webb Space Telescope. The data were obtained from the Mikulski Archive for STScI. These observations are associated with program \# 8047, 8204.
Some of the data products presented herein were retrieved from the Dawn JWST Archive (DJA). DJA is an initiative of the Cosmic Dawn Center (DAWN), which is funded by the Danish National Research Foundation under grant DNRF140.
This publication made use of the NASA Astrophysical Data System for bibliographic information.

\end{acknowledgments}

\facilities{JWST (NIRSpec)}

\software{
  Astropy \citep{2013A&A...558A..33A, 2018AJ....156..123A, 2022ApJ...935..167A},
  Matplotlib \citep{2007CSE.....9...90H},
  NumPy \citep{2020Natur.585..357H},
  NumPyro \citep{2019arXiv191211554P, Pyro}
}

\appendix

\section{Gaussian versus Exponential Profiles for the Non-Balmer Broad Components\label{sec:app:expnb}}

In our fiducial model the broad Balmer emission is exponential, whereas the broad components of the non-Balmer lines (\neiiiw, \oiiiauroral, \oiiidoubb, and \hei\,$\lambda5877,\lambda7067$) are Gaussian (Section~\ref{sec:method:fit}). Because several of these non-Balmer broad components are reported here for the first time, we test whether the data prefer either shape by refitting both sources with the non-Balmer broad components switched to an exponential profile. The swap is parameter-count neutral, with a single width parameter in either case, so the fits are directly comparable.

\begin{figure*}
\gridline{
  \fig{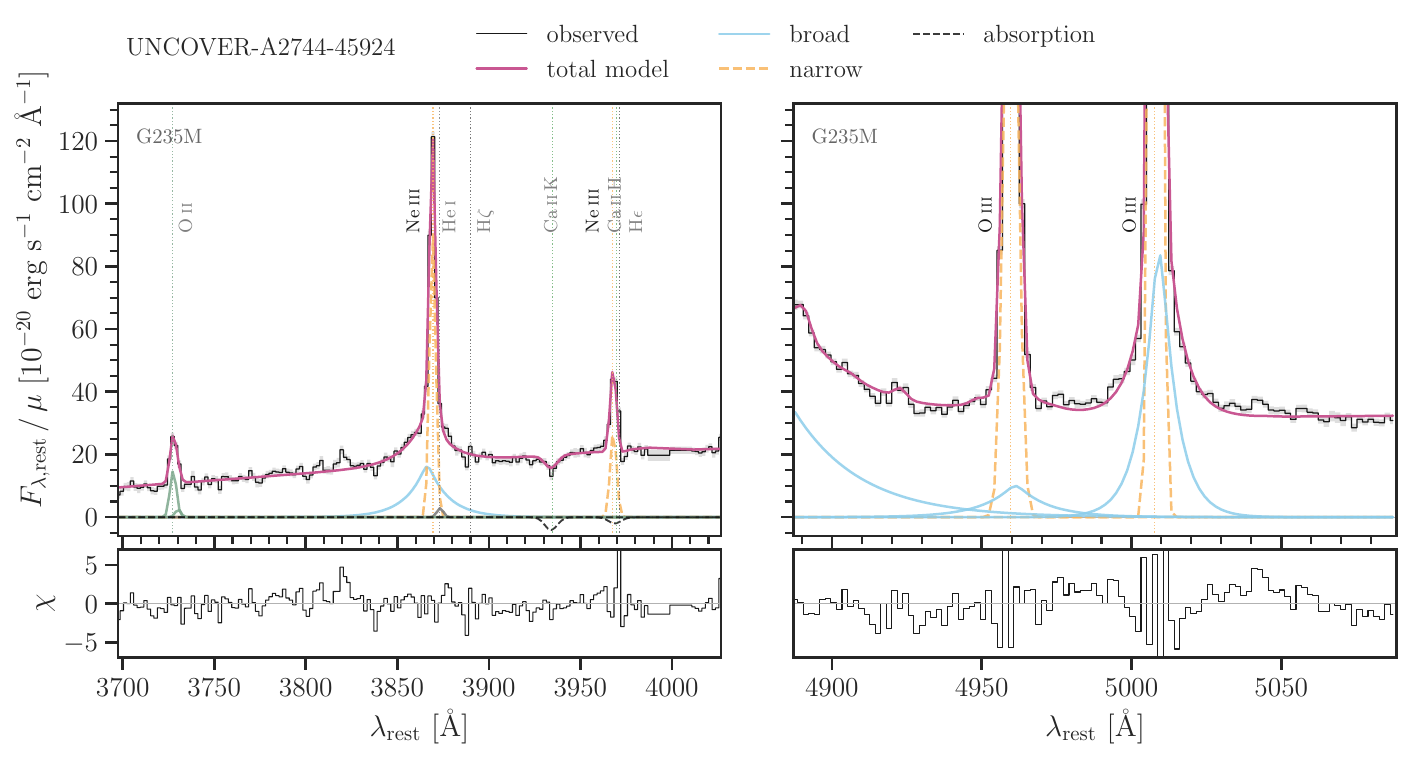}{0.95\textwidth}{(a)}
}
\gridline{
  \fig{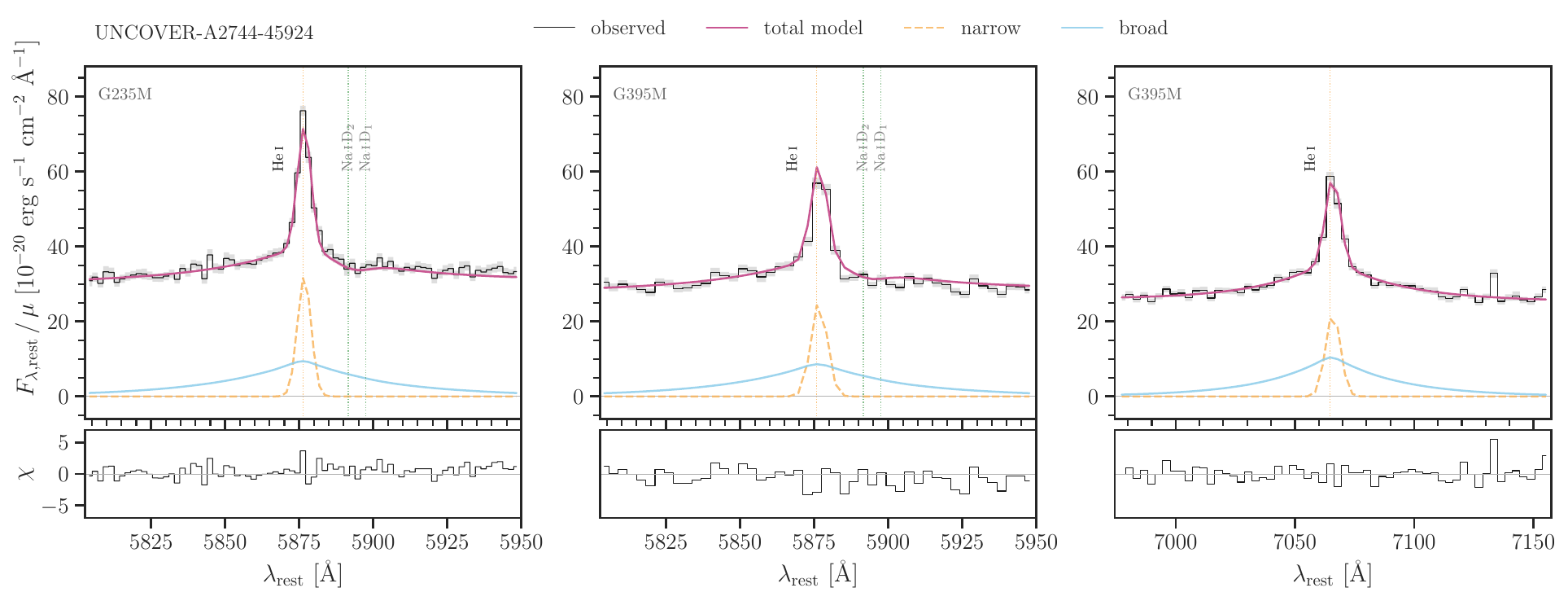}{0.95\textwidth}{(c)}
}
\caption{Exponential-profile fits to the broad non-Balmer components in \ms:
  (a) \neiiiw, and \oiiidoubb, and (b) the \heib\ complex. These are
  statistically indistinguishable from the fiducial Gaussian fits.
}
\label{fig:expnb_monster}
\end{figure*}

\begin{figure*}
\gridline{
  \fig{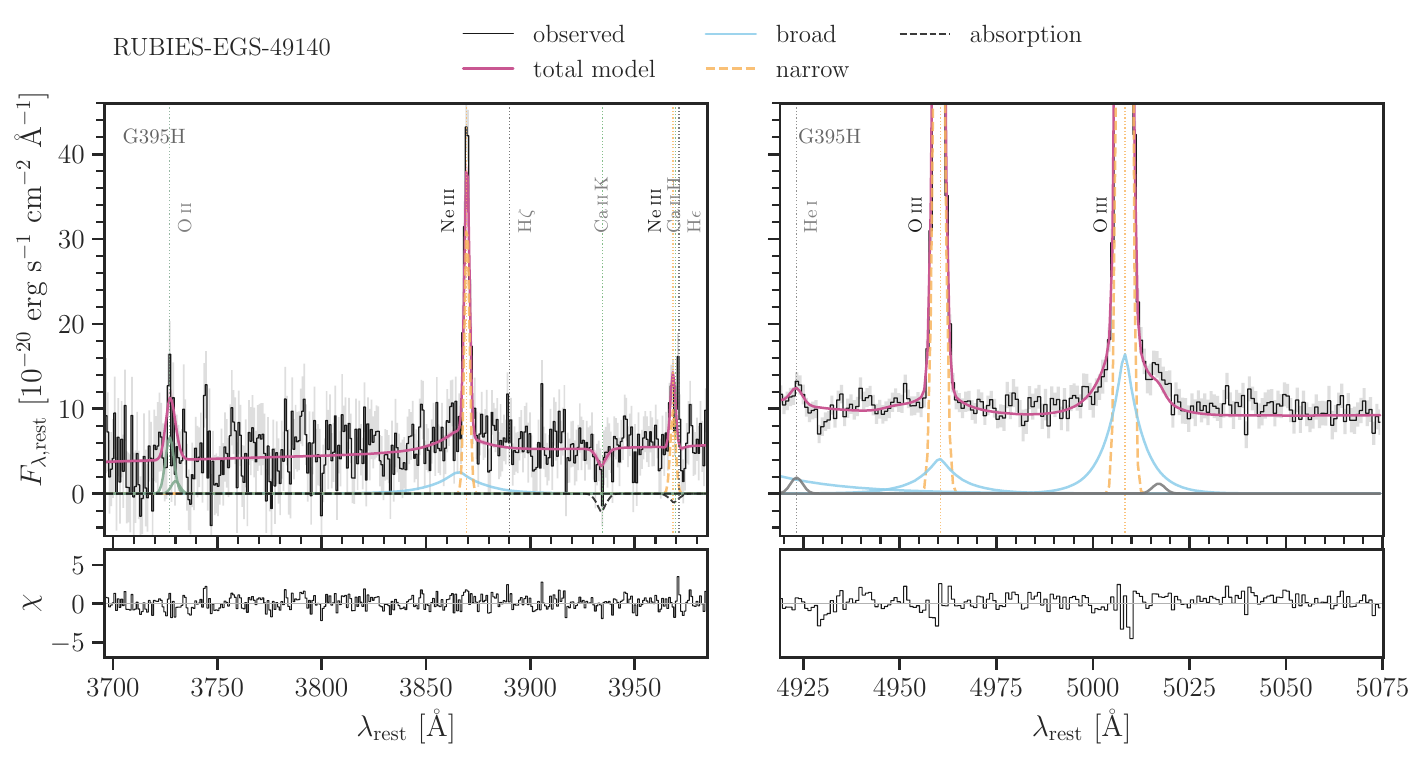}{0.95\textwidth}{}
}
\caption{Exponential-profile fits to the broad components of \neiiiw\ (left) and \oiiidoubb\ (right) in \rd.
}
\label{fig:expnb_49140}
\end{figure*}

The exponential-profile fits are shown in Figures~\ref{fig:expnb_monster} and \ref{fig:expnb_49140} for \ms\ and \rd, respectively. We quantify the comparison using the Watanabe-Akaike information criterion (WAIC). Neither functional form is preferred in either source. The intrinsic shape of the broad non-Balmer emission therefore remains unconstrained by the current data, despite the exceptional S/N. Further progress will therefore more likely come from physically motivated modeling to guide the data models on whether the broad non-Balmer emission arises from a distinct component.

\section{The [Fe\,{\sc vii}]$\,\lambda$5160 vs. [Fe\,{\sc ii}]\,$\lambda$5159 Identification\label{sec:app:fevii}}

\begin{figure*}
\gridline{
  \fig{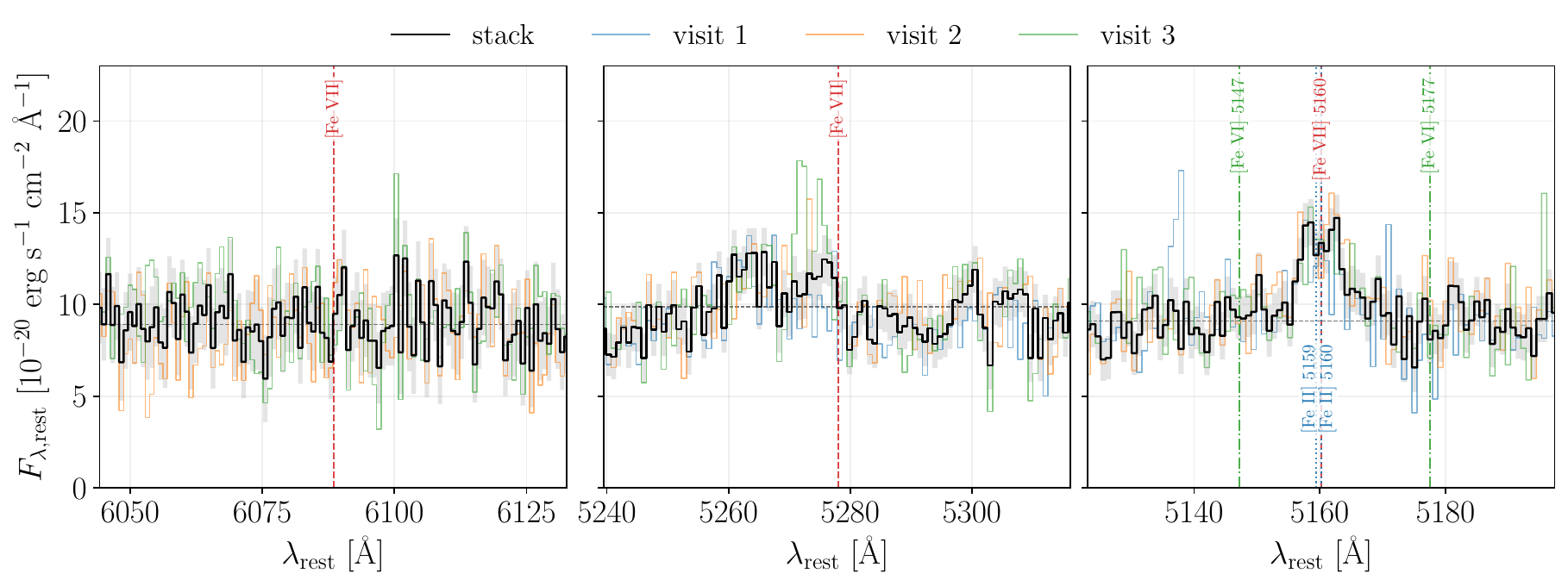}{0.95\textwidth}{}
}
\caption{The [Fe\,{\sc vii}]$\lambda$5160 versus [Fe\,{\sc ii}]$\lambda$5159 identification for \rd. (Left) [Fe\,{\sc vii}]$\lambda$6088.56 is the strongest expected transition in the multiplet group (multiplet 0). (Middle) [Fe\,{\sc vii}]$\lambda$5278.02 (multiplet 1) is predicted to be $\sim 1.77\times$ stronger than [Fe\,{\sc vii}]$\lambda$5160. (Right) The debated [Fe\,{\sc vii}]$\lambda$5160 feature, which is blended with [Fe\,{\sc ii}]$\lambda$5160.
}
\label{fig:fevii_test}
\end{figure*}

A detection of [Fe\,{\sc vii}]\,$\lambda$5160 in \rd\ was reported by \citet{Lambrides2025:fevii}, which \citet{DEugenio2025:fe} later identified as [Fe\,{\sc ii}]\,$\lambda$5159. Both identifications were based on G395M spectroscopy; we therefore revisit this region using our higher-resolution spectra (Figure~\ref{fig:fevii_test}). Unfortunately, [Fe\,{\sc ii}] transitions can fall at the same wavelength as [Fe\,{\sc vii}]\,$\lambda$5160, so the feature alone cannot distinguish between the two identifications.
A closer investigation of possible velocity offsets between the different transitions could provide additional constraints on, and potentially support, the [Fe\,{\sc vii}] identification. However, interpreting Fe emission is non-trivial (e.g., \citealt{Boroson1992, Laor1997, Sigut1998, VeronCetty2004, Bruhweiler2008, Kovacevic2010, Park2022}), and a detailed analysis is beyond the scope of this work.

We further search for additional [Fe\,{\sc vii}] transitions. In particular, [Fe\,{\sc vii}]\,$\lambda$6088 and $\lambda$5278 are expected to be stronger than $\lambda$5160, yet neither is detected. Moreover, the weakness of the \heiir\ line \citep{Wang2026:qion} argues against the extreme ionization conditions required to produce [Fe\,{\sc vii}] emission (ionization potential $=99$~eV). A radiation field capable of generating detectable [Fe\,{\sc vii}]\,$\lambda$5160 would be expected to produce substantially stronger \heiir. The absence of the stronger [Fe\,{\sc vii}] transitions, as also pointed out by \citet{DEugenio2025:fe}, together with the weak \heiir\ emission, therefore seems to disfavor the [Fe\,{\sc vii}]\,$\lambda$5160 interpretation.

\bibliography{lrd_g395h_wang.bib}
\bibliographystyle{aasjournal}

\end{CJK*}
\end{document}